\documentclass{aa}
\usepackage[draft]{hyperref}
\usepackage[varg]{txfonts}

\bibpunct{(}{)}{;}{a}{}{,} 

\begin{document}

%
%
    \title{All-sky census of Galactic high-latitude molecular intermediate-velocity clouds}
    \titlerunning{High-latitude molecular intermediate-velocity clouds}


   \author{T.~R{\"o}hser \inst{1} 
           \and
           J.~Kerp \inst{1}
           \and 
           D.~Lenz \inst{1}
           \and
           B.~Winkel \inst{2}
           }

   \institute{Argelander-Institut f{\"u}r Astronomie (AIfA), Universit{\"a}t Bonn, Auf dem H{\"u}gel 71, D-53121 Bonn\\
              \email{troehser@astro.uni-bonn.de}
              \and
              Max-Planck-Institut f{\"u}r Radioastronomie (MPIfR), Auf dem H{\"u}gel 69, D-53121 Bonn
              }

   \date{Received ---; accepted ---}

   \abstract
   {The \ion{H}{i} halo clouds of the Milky Way, and in particular the intermediate-velocity clouds (IVCs), are thought to be connected to Galactic fountain processes. Observations of fountain clouds are important for understanding the role of matter recycling and accretion onto the Galactic disk and subsequent star formation.}
   {Here, we quantify the amount of molecular gas in the Galactic halo. We focus on the rare class of molecular IVCs (MIVCs) and search for new objects.}
   {The \ion{H}{i}-FIR correlation was studied across the entire northern and southern Galactic hemispheres at Galactic latitudes $|b|>20^\circ$ to determine the amount and distribution of molecular gas in IVCs. We used the most recent large-scale \ion{H}{i} and FIR data, the Effelsberg Bonn-\ion{H}{i} Survey, the Parkes Galactic All-Sky Survey, and the \textit{Planck} FIR surveys.}
   {We present a catalogue of 239 MIVC candidates on the northern and southern Galactic hemispheres. Among these candidates, all previously known MIVCs are recovered except for one single source. The frequency of candidates differs significantly between the northern and southern Galactic hemispheres and between negative and positive LSR velocities as well.}
   {In our approach we analyse the local Galactic environment. Extrapolating our results to the entire Galaxy, the global inflow of atomic and molecular IVC gas onto the Milky Way may account for the major fraction of the gaseous mass that is required to sustain the current Galactic star formation rate.}


   \maketitle

\section{Introduction}
\label{sec:introduction}

A star-forming galaxy such as the Milky Way requires the continuous accretion of matter to sustain its star formation activity over timescales of Gyr \citep[e.g.][]{Fraternali2012,Putman2012}. The accreted material needs to be of low metallicity to match the observed abundances in Galactic stellar populations, which is known as the G-dwarf problem \citep{Alibes2001}. Gaseous halos of galaxies are thought to be an important reservoir for such material \citep[e.g.][]{Wakker2001}.

The intermediate- and high-velocity clouds (IVCs and HVCs) are prominent structures of extra-planar cold atomic gas in our Galaxy \citep[][for a recent review on gaseous halos]{Putman2012}. These classes of \ion{H}{i} halo clouds are defined by their anomalous radial velocity that is incompatible with simple rotation models. IVCs are typically defined with $40\,\mathrm{km}\,\mathrm{s}^{-1}\leq |v_\mathrm{LSR}| \leq 90\,\mathrm{km}\,\mathrm{s}^{-1}$, and HVCs with $90\,\mathrm{km}\,\mathrm{s}^{-1}\leq |v_\mathrm{LSR}|$ \citep[e.g.][]{Wakker2001}.

The IVCs are located within a distance of $\sim$2\,kpc and consist of gas of near solar metallicity. The HVCs have distances of $5\lesssim D \lesssim 20\,$kpc and lower metallicities \citep[e.g.][]{Wakker2001}. A main exception is the Magellanic System, which is located at much larger distance \citep[e.g.][]{Nidever2010}. Commonly, FIR emission is detected towards IVCs \citep{Planckcollaboration2011XXIV}, however, no dust emission could be convincingly associated with HVCs so far \citep{Wakker1986,Planckcollaboration2011XXIV,Williams2012,Saul2014,Lenz2016}, and in other cases they are only tentatively detected \citep{MivilleDeschenes2005a,Peek2009}. These properties suggest that IVCs may be connected to a Galactic fountain process, while HVCs represent an extragalactic reservoir of gas \citep[e.g.][]{Putman2012}.

According to this picture, IVCs are considered as an inflow of gas consisting predominantly of recycled disk material. During the fountain cycle, clouds may gain mass by the local cooling of the hot halo medium \citep{Marinacci2010,Marasco2012}. As fountain clouds, IVCs may therefore contribute to the global accretion.

Star formation requires cold and dense gas as fuel \citep[e.g.][]{Putman2012}. Such gas may harbour molecules, whose presence can be used to identify cold and dense gas. Studies using absorption spectroscopy reveal a widespread distribution of low molecular hydrogen (H$_2$) column densities of $N_{\mathrm{H}_2}=10^{14}-10^{16}\,\mathrm{cm}^{-2}$ within IVCs \citep{Richter2003,Wakker2006}.

There are a few molecular intermediate-velocity clouds (MIVCs) that contain significant amounts of molecular gas, such that carbon monoxide $^{12}$CO(1$\rightarrow$0) emission is detected \citep{Magnani2010}. These MIVCs are of prime interest for studies of inflowing cold, dense, and molecular material onto the Milky Way \citep[e.g.][]{MivilleDeschenes2016}. MIVCs are located at scale heights of a few hundred parsecs above the disk with in-falling radial velocities \citep{Magnani2010}. These rare objects are Draco, IVC\,135, IVC\,210, G\,283.8+54.9, G\,288.4+53.2, and G\,295.0+57.1. The latter three appear to be part of the same gaseous structure \citep{Magnani2010}. Because of their large amounts of molecular gas, Draco \citep{Mebold1985,Mebold1987,Rohlfs1989,Herbstmeier1993,Herbstmeier1996,Moritz1998,Heithausen2001b,MivilleDeschenes2016}, IVC\,135 \citep{Heiles1988,Weiss1999,Heithausen2001b,Hernandez2013,Lenz2015}, and IVC\,210 \citep{Desert1990a,Roehser2014,Roehser2016} have been extensively studied. Furthermore, there are confirmed distance limits for these three MIVCs \citep{Gladders1998,Benjamin1996,Roehser2014}.

The low number of MIVCs suggests that either MIVCs are rare and require special conditions to form, or that only a small fraction of the MIVC population has been detected so far. The CO survey conducted by \citet{Dame2001} covers only latitudes of $|b|\lesssim20^\circ$ , except for a few known high-latitude molecular structures. The CO maps extracted from the \textit{Planck} surveys \citep{Planckcollaboration2014XIII} only have a limited sensitivity and do not provide spectral information.

We have therefore to rely on the indirect inference of molecular hydrogen on Galactic scales. One method is the correlation of \ion{H}{i} and FIR dust emission and the analysis of its residuals. With respect to the linear \ion{H}{i}-FIR correlation, FIR excess is inferred, which is commonly attributed to the presence of molecular hydrogen \citep{Desert1988,Reach1994,Reach1998}. 

The most recent and comprehensive study of FIR-excess emission was conducted by \citet{Reach1998}. However, they only considered the total \ion{H}{i} column density within $-100\,\mathrm{km}\,\mathrm{s}^{-1}\leq v_\mathrm{LSR} \leq 100\,\mathrm{km}\,\mathrm{s}^{-1}$ and therefore did not make use of the velocity information of the \ion{H}{i} data. In addition, the angular resolution and quality of their data, especially in \ion{H}{i}, is significantly surpassed by the new all-sky surveys that are available today \citep{Winkel2016,Kalberla2015,Planckcollaboration2013I}.

Here, we present an all-sky search for MIVCs focussing  on high Galactic latitudes of $|b|>20^\circ$. At lower latitudes many different gas components are superimposed along the lines of
sight, which does not allow reliably distinguishing different FIR emission components by their \ion{H}{i} counterparts. 

This paper is organised as follows. In Sect.~\ref{sec:data} we present the \ion{H}{i} and FIR data that we use in this all-sky analysis. In Sect.~\ref{sec:methods} we discuss the \ion{H}{i}-FIR correlation and how we estimate H$_2$ column densities on Galactic scales. In addition, MIVC candidates are defined by a set of observed quantities. In Sect.~\ref{sec:nor-gal-hem} we present the distribution of the inferred MIVC candidates on the northern Galactic hemisphere at $b>20^\circ$. In Sect.~\ref{sec:sou-gal-hem} the same analysis is shown for the southern Galactic hemisphere at $b<-20^\circ$. An additional analysis of the global MIVC samples is presented in Sect.~\ref{sec:results}. Our results are discussed in Sect.~\ref{sec:discussion} and we conclude in Sect.~\ref{sec:summary}.

\section{Data}
\label{sec:data}

\subsection{\texorpdfstring{\ion{H}{i}}{HI} data}
\label{sec:data-hi}

We combined the most recent large-scale \ion{H}{i} surveys to obtain an all-sky coverage for the atomic hydrogen. On the northern hemisphere the Effelsberg-Bonn \ion{H}{i} Survey is available \citep[EBHIS,][]{Winkel2016,Kerp2011,Winkel2010}, which covers the sky at declinations of $\delta\geq-5^\circ$ with the Effelsberg 100\,m telescope. The data have a half-power beam width of $\theta\simeq10.8\arcmin$ and a spectral channel width of $\Delta v \simeq 1.29\,\mathrm{km}\,\mathrm{s}^{-1}$ (Table \ref{tab:data}). 

The Parkes Galactic All-Sky Survey \citep[GASS,][]{McClureGriffiths2009,Kalberla2010,Kalberla2015} is a fully sampled survey of Galactic \ion{H}{i} with the Parkes 64\,m telescope for declinations of $\delta\leq1^\circ$. The angular resolution of the data is $\theta\simeq16.1\arcmin$ with a channel spacing of $\Delta v\simeq0.82\,\mathrm{km}\,\mathrm{s}^{-1}$ (Table \ref{tab:data}).

Both EBHIS and GASS are corrected for stray radiation, which is an emission component that enters the receiving system of the telescopes through their side lobes \citep{Kalberla1980}. The relative contribution to the measured signal is largest at high Galactic latitudes and therefore has to be removed \citep[e.g.][their Fig.~A.2.]{Winkel2016}.

These two surveys were combined into a single all-sky \ion{H}{i} data set. To achieve a uniform coverage, we smoothed the EBHIS data to the angular resolution of GASS. Furthermore, the GASS data were smoothed to the spectral resolution of EBHIS. In the overlap region of EBHIS and GASS at $-5^\circ\leq\delta\leq1^\circ$ , the two surveys are averaged. The FITS\footnote{http://fits.gsfc.nasa.gov/} maps were computed in zenith-equal area projection \citep{Calabretta2002} using the convolution-based gridder \textit{cygrid} \citep{Winkel2016b}.

Additionally, we decomposed the EBHIS and GASS data cubes into Gaussian components. This decomposition was achieved by simultaneously fitting a varying number of Gaussian components to each individual spectrum. The Akaike information criterion \citep{Akaike1974} was used to avoid overfitting by penalising models with more parameters. At most 15 components per spectrum were fitted. The amplitudes were required to be positive and larger than the noise, and the line widths had to reflect reasonable gas temperatures for the interstellar medium (ISM). These Gaussian components were used to identify cold atomic gas (Sect.~\ref{sec:def-mivc}). The Gaussian decomposition is not unique. However, narrow and bright line components are identified easily because they dominate the \ion{H}{i} emission along their lines of sight.

\begin{table}
  \caption{Data sets used in this study. The columns list the angular resolution $\theta$, the spectral channel width $\Delta v$, and the brightness temperature noise $\Delta T$.}  
  \label{tab:data}
  \small
  \centering
  \begin{tabular}{ccccc}
    \hline\hline
     Data & $\theta$ & $\Delta v$ & $\Delta T$ & References \\
     & [arcmin] & [km\,s$^{-1}$] & [mK] & \\
     \hline
      EBHIS & 10.8 & 1.29 & 90 & (1) \\
      GASS & 16.1 & 0.82 & 57 & (2)\\
      \textit{Planck} $\tau$ & 5.27 & -- & -- & (3) \\      
     \hline
  \end{tabular}
  \tablebib{
  (1) \citet{Winkel2016,Kerp2011,Winkel2010}; (2) \citet{McClureGriffiths2009,Kalberla2010,Kalberla2015}; (3) \citet{Planckcollaboration2014XI}.
  }
\end{table}

\subsection{FIR data}
\label{sec:data-fir}

The FIR intensity of dust continuum emission can be described by the spectrum of a modified black-body \citep[e.g.][]{Boulanger1996}. Dust optical depths $\tau$ were inferred by \citet{Planckcollaboration2014XI} from data measured by the \textit{Planck} satellite at 353\,GHz, 545\,GHz, and 857\,GHz in addition to the 3000\,GHz data from IRIS \citep{MivilleDeschenes2005b}. The dust models are provided on a HEALPix grid \citep{Gorski2005} at an angular resolution of $5\arcmin$. Owing to the gridding to FITS maps, the angular resolution is slightly degraded (Table \ref{tab:data}). The dust optical depths $\tau$ were smoothed to the angular resolution of GASS for the \ion{H}{i}-$\tau$ correlation study and the inference of $H_2$ column densities. In addition, the point source mask was applied that has been derived by \citet{Planckcollaboration2015XXVI} at 857\,GHz.

\section{Methods}
\label{sec:methods}

A linear correlation between the total hydrogen column density and the FIR dust emission exists
within the diffuse ISM \citep[e.g.][]{Boulanger1996}. This linear correlation allows the removal of the FIR emission that is associated with the atomic phase \citep{Desert1988,Reach1998}. The residual emission contains many different contributions, such as~intrinsic variations of the \ion{H}{i}-FIR correlation or variations of the gas-to-dust ratio, or dust that is associated with ionised or molecular hydrogen \citep[e.g.][]{Planckcollaboration2011XIX}. All these effects may cause structured residual FIR emission.

\subsection{H-FIR correlation}
\label{sec:h-fir-correlation}

The following equations describe the general correlation between gas and dust in the ISM at Galactic coordinates $(l,b)$ \citep[e.g.][]{Lenz2015,Roehser2016}:
\begin{align}
 \label{eq:h-ir-corr}
 \tau(l,b) &= R + \epsilon \times N_\mathrm{H}(l,b) \nonumber\\ 
 &\simeq R + \epsilon \times \left[N_\mathrm{\ion{H}{ii}}(l,b) + N_\mathrm{\ion{H}{i}}(l,b) + 2N_{\mathrm{H}_2}(l,b)\right].
\end{align}
There are two linear parameters in this representation, the constant offset $R,$ and the dust emissivity $\epsilon$. Since we focus here on the inference of molecular hydrogen in IVCs, we considered clouds with \ion{H}{i} column densities $N_\mathrm{\ion{H}{i}}\geq1\times10^{20}\,\mathrm{cm}^{-2}$, for which the transition from atomic to molecular hydrogen is observed to occur \citep[e.g.][]{Gillmon2006a}. In such environments the ionised hydrogen column density is thought to be negligible \citep{Lagache2000}. The assumption of a single dust emissivity for the atomic and molecular gas is a simplification that ignores detected small variations of dust emission properties even in the diffuse ISM \citep{Ysard2015,Fanciullo2015}.

The kinematical information on the \ion{H}{i} gas was used to distinguish between different gaseous components: low-velocity, intermediate-velocity, and high-velocity clouds (LVCs, IVCs, and HVCs). For the LVCs we chose $-20\,\mathrm{km}\,\mathrm{s}^{-1}\leq v_\mathrm{LSR} \leq +20\,\mathrm{km}\,\mathrm{s}^{-1}$, for the IVCs $20\,\mathrm{km}\,\mathrm{s}^{-1}\leq |v_\mathrm{LSR}| \leq 100\,\mathrm{km}\,\mathrm{s}^{-1}$, and for the HVCs $100\,\mathrm{km}\,\mathrm{s}^{-1}\leq |v_\mathrm{LSR}|\leq450\,\mathrm{km}\,\mathrm{s}^{-1}$. The velocity limit for the distinction between LVC and IVC gas is rather low \citep[e.g.][]{Wakker2001}. However, some of the known high-latitude molecular clouds have radial velocities of about $-25\,\mathrm{km}\,\mathrm{s}^{-1}$, like Draco, G\,288.4+53.2, or G\,295.0+57.1 \citep{Magnani2010}. These objects were included in our IVC definition.

We calculated all-sky \ion{H}{i} column density maps of the LVC gas for $-20\,\mathrm{km}\,\mathrm{s}^{-1}\leq v_\mathrm{LSR} \leq +20\,\mathrm{km}\,\mathrm{s}^{-1}$, negative IVC gas $N_\mathrm{\ion{H}{i}}^{\mathrm{IVC}^-}$ with $-100\,\mathrm{km}\,\mathrm{s}^{-1}\leq v_\mathrm{LSR} \leq -20\,\mathrm{km}\,\mathrm{s}^{-1}$, and positive IVC gas $N_\mathrm{\ion{H}{i}}^{\mathrm{IVC}^+}$ at $+20\,\mathrm{km}\,\mathrm{s}^{-1}\leq v_\mathrm{LSR} \leq +100\,\mathrm{km}\,\mathrm{s}^{-1}$. 

The HVC \ion{H}{i} gas is neglected in the \ion{H}{i}-$\tau$ correlation since it is not associated with detectable amounts of FIR emission \citep[e.g.][]{Planckcollaboration2011XXIV,Lenz2016}. The observed dust optical depths $\tau$ can be described by a two-component model
\begin{equation}
 \label{eq:h-ir-corr2}
  \tau \simeq R + \epsilon^{\mathrm{LVC}} \times (N_\mathrm{\ion{H}{i}}^{\mathrm{LVC}} + 2N_{\mathrm{H}_2}^{\mathrm{LVC}}) + \epsilon^{\mathrm{IVC}} \times (N_\mathrm{\ion{H}{i}}^{\mathrm{IVC}} + 2N_{\mathrm{H}_2}^{\mathrm{IVC}})
\end{equation}
that distinguishes between LVCs and IVCs. Both components may have different gas-to-dust ratios expressed as different emissivities $\epsilon^i$ \citep[e.g.][]{Planckcollaboration2011XXIV}. When there is FIR excess emission or molecular hydrogen associated with LVCs, these lines of sight are difficult to analyse since we cannot distinguish uniquely between excess emission that originates from LVC or IVC gas. Hence, we assumed $N_{\mathrm{H}_2}^{\mathrm{LVC}}\simeq0$, assigning the entire excess emission to the IVC component. This conclusion is incorrect, however, when significant gas and dust emission is associated with the LVC regime. Consequently, these regions were excluded from the further analysis to a large extent in our definition of MIVCs (Sect.~\ref{sec:def-mivc}).

Rearranging Eq.~\eqref{eq:h-ir-corr2} yields
\begin{equation}
 \label{eq:nh2-ivc}
  N_{\mathrm{H}_{2}}^{\mathrm{IVC}} \simeq \frac{1}{2} \left( \frac{\tau-(R+\epsilon^{\mathrm{LVC}}\times N_{\mathrm{\ion{H}{i}}}^{\mathrm{LVC}})}{\epsilon^{\mathrm{IVC}}}-N_{\mathrm{\ion{H}{i}}}^{\mathrm{IVC}} \right),
\end{equation}
allowing us to estimate H$_2$ column densities within IVCs across large angular scales.

\subsection{Fitting the \texorpdfstring{\ion{H}{i}}{HI}-\texorpdfstring{$\tau$}{t} correlation}
\label{sec:fit-hi-tau}

The parameters $R$, $\epsilon^\mathrm{LVC}$, and $\epsilon^\mathrm{IVC}$ were obtained from fits of Eq.~\eqref{eq:h-ir-corr2} to the \textit{\textup{linear}} part of the H-$\tau$ correlation, which is assumed to be dominated by atomic hydrogen. However, at some threshold \ion{H}{i} column density, the atomic hydrogen turns molecular, which introduces the non-linear FIR excess that steepens the linear \ion{H}{i}-$\tau$ correlation \citep{Desert1988}. The \ion{H}{i}-H$_2$ transition is smooth and continuous (compare with Fig.~\ref{fig:hi-tau-north}).

To quantify the non-linear contribution of H$_2$ to the H-$\tau$ correlation, we applied the iterative fitting method as described by \citet{Roehser2016}. The two-component model (Eq.~\ref{eq:h-ir-corr2}) was fitted below an upper \ion{H}{i} column density threshold of $4\times10^{20}\,\mathrm{cm}^{-2}$ in either LVC or IVC \ion{H}{i} column density.

We fitted the two-component model separately for the entire northern and southern Galactic hemispheres with $b>20^\circ$ or $b<-20^\circ$ to determine the average Galactic gas-to-dust ratios. The analysis was performed separately for the two Galactic hemispheres mainly because of the Magellanic System \citep[e.g.][]{Nidever2010}, which may affect the correlation results (Sect.~\ref{sec:sou-gal-hem}).

\subsection{Definition of MIVCs}
\label{sec:def-mivc}

\begin{table*}
  \caption{Fitted global linear parameters for the northern and southern Galactic hemispheres at latitudes of $|b|>20^\circ$. Details of the fitting procedure are given in the text. The statistical errors on the parameters are of the order of $\sim$$10^{-4}$ in the given units.}  
  \label{tab:hi-tau-coeff}
  \footnotesize
  \centering
  \begin{tabular}{ccc|ccc}
    \hline\hline
    \multicolumn{3}{c}{north} & \multicolumn{3}{c}{south} \\
    \hline
    $R$ & $\epsilon^\mathrm{LVC}$ & $\epsilon^\mathrm{IVC}$ & $R$ & $\epsilon^\mathrm{LVC}$ & $\epsilon^\mathrm{IVC}$\\
    $\left[10^{-6}\right]$ & $\left[10^{-26}\,\mathrm{cm}^{2}\right]$ & $\left[10^{-26}\,\mathrm{cm}^{2}\right]$ & $\left[10^{-6}\right]$ & $\left[10^{-26}\,\mathrm{cm}^{2}\right]$ & $\left[10^{-26}\,\mathrm{cm}^{2}\right]$\\
    \hline
    $0.068$ & $0.754$ & $0.451$ & $-0.148$ & $0.723$ & $0.637$ \\      
    \hline
  \end{tabular}
\end{table*}

The Gaussian decomposition of the GASS reveals that the Galactic atomic hydrogen is characterised by narrow and wide spectral lines that correspond to cold and warm gas \citep[][their Fig.~1]{Kalberla2015}. The cold atomic component may be associated with molecular gas \citep[e.g.][]{Snow2006}. Accordingly, we defined selection criteria for MIVCs that target cold \ion{H}{i} gas with high line intensities.

These criteria match the observable properties of the well-studied MIVCs IVC\,135 \citep{Weiss1999} and IVC\,210 \citep{Desert1990a}. These two objects have been extensively studied with data from the Effelsberg 100\,m telescope \citep{Weiss1999,Lenz2015,Roehser2014}. Both IVC\,135 and IVC\,210 have peak \ion{H}{i} column densities of $N_\mathrm{\ion{H}{i}}^\mathrm{peak}\simeq2.5-3\times10^{20}\,\mathrm{cm}^{-2}$ and a line width of the cold \ion{H}{i} gas with a $\mathrm{FWHM}\simeq4-5\,\mathrm{km}\,\mathrm{s}^{-1}$. These objects are well known because they dominate the total gas and dust emission along their lines of sight. An MIVC observationally contains 1) a sufficiently high IVC \ion{H}{i} column density, 2) cold gas within the IVC velocity regime $20\,\mathrm{km}\,\mathrm{s}^{-1}\leq |v_\mathrm{LSR}| \leq 100\,\mathrm{km}\,\mathrm{s}^{-1}$, and 3) a statistically significant H$_2$ column density. In addition, there is 4) only little LVC gas.
\begin{enumerate}
 \item For IVC\,135 and IVC\,210 the transition from atomic to molecular hydrogen occurs at $N_\mathrm{\ion{H}{i}}^0=1-2\times10^{20}\,\mathrm{cm}^{-2}$ \citep{Weiss1999,Lenz2015,Roehser2014,Roehser2016}. We required the MIVC candidates to have $N_\mathrm{\ion{H}{i}}^\mathrm{IVC}>1\times10^{20}\,\mathrm{cm}^{-2}$.
 \item Rapid H$_2$ formation is expected to be associated with cold and dense gas \citep[e.g.][]{Snow2006}. The \ion{H}{i} line widths set upper limits on the kinetic temperature of the gas. Observationally, MIVCs have Gaussian line components with amplitudes larger than A$_\mathrm{gauss}>10\,$K and line widths of FWHM$_\mathrm{gauss}<5\,\mathrm{km}\,\mathrm{s}^{-1}$, corresponding to $T_\mathrm{kin}<545\,$K. This is a mild upper limit since the actual gas temperatures are significantly lower, as is inferred by CO observations \citep[e.g.][]{Heyer2015}. The line parameters were not extracted from the \ion{H}{i} survey data directly but from a Gaussian decomposition (Sect.~\ref{sec:data-hi}).
 \item The intrinsic scatter of the \ion{H}{i}-FIR correlation was used as a measure for the statistical significance of the derived H$_2$ column densities $N_{\mathrm{H}_2}$. Towards regions with $N_\mathrm{\ion{H}{i}}^\mathrm{TOT}=N_\mathrm{\ion{H}{i}}^\mathrm{LVC}+N_\mathrm{\ion{H}{i}}^\mathrm{IVC}<1\times10^{20}\,\mathrm{cm}^{-2}$ , no H$_2$ is expected and the scatter of the \ion{H}{i}-FIR correlation is assumed to be caused by intrinsic variations of the correlation. For these parts of the sky, we calculated the median $\tilde{N}_{\mathrm{H}_2}$ and standard deviation $\sigma(N_{\mathrm{H}_2})$ across the hemispheres  from
the derived H$_2$ map (Sect.~\ref{sec:h-fir-correlation}). This translates into $3\sigma$ criteria of $N_{\mathrm{H}_2}>\tilde{N}_{\mathrm{H}_2}+3\sigma(N_{\mathrm{H}_2})$.
 \item When the LVC \ion{H}{i} column density towards IVC gas is high, molecules may be located within LVCs, which causes FIR excess emission that might incorrectly be associated with IVC gas. Furthermore, the imprint of the IVC on the \ion{H}{i}-$\tau$ correlation may be marginal and, thus, FIR excess within the IVC gas may not be correctly identified. We therefore required MIVC candidates to significantly contribute to the total \ion{H}{i} column density along their lines of sight by choosing $N_\mathrm{\ion{H}{i}}^\mathrm{LVC}<2N_\mathrm{\ion{H}{i}}^\mathrm{IVC}$.
\end{enumerate}
More candidates can be obtained by relaxing either criterion. Our goal here is, however, to find all of the clear and distinct MIVC candidates.

We identified extended regions that fulfilled these four criteria simultaneously. We did not require the candidates to have a minimum angular size since there may be only a few connected pixels that satisfy all criteria and exceed the required thresholds. Nevertheless, these candidates are also interesting objects since they may indicate the most obvious manifestation of a potential MIVC. In the subsequent analysis we distinguish between objects that have angular extents that are larger or smaller than the resolution element of our study, which is a single GASS beam (Table \ref{tab:data}).

To allow small angular offsets between the different observables ($N_\mathrm{\ion{H}{i}}^\mathrm{IVC}$, A$_\mathrm{gauss}$, FWHM$_\mathrm{gauss}$, $N_{\mathrm{H}_2}^\mathrm{IVC}$, $N_\mathrm{\ion{H}{i}}^\mathrm{LVC}$), we applied maximum filters with the size of the angular resolution of GASS. Small shifts between the different observables are likely to occur, for instance because of the FIR emission of LVC gas.

Our search distinguishes between IVC gas at negative and positive LSR velocities with $-100\,\mathrm{km}\,\mathrm{s}^{-1}\leq v_\mathrm{LSR} \leq -20\,\mathrm{km}\,\mathrm{s}^{-1}$ or $+20\,\mathrm{km}\,\mathrm{s}^{-1}\leq v_\mathrm{LSR} \leq +100\,\mathrm{km}\,\mathrm{s}^{-1}$. In the following these two different regimes are denoted as MIVC$^\pm$.

\subsection{Ranking of inferred MIVC candidates}
\label{sec:ranking}

There is a large number of inferred MIVC candidates, some of which are more reliable than others. We quantified the significance $S$ of each candidate by
\begin{equation}
\label{eq:significance}
 S = \frac{\epsilon^\mathrm{IVC}_{\mathrm{cand}}}{\sigma \left(\epsilon^\mathrm{IVC}_{\mathrm{cand}}\right)}\times\sqrt{N}
,\end{equation}
where $\epsilon^\mathrm{IVC}_\mathrm{cand}$ is the estimated slope of the \textit{local} \ion{H}{i}-$\tau$ correlation in a small region of $2^\circ\times2^\circ$ around the centre of each candidate. Moreover, $\sigma \left(\epsilon^\mathrm{IVC}_\mathrm{cand}\right)$ is the corresponding fitting uncertainty and $N$ is the number of pixels of each candidate, enhancing larger regions that fulfil the criteria. The local \ion{H}{i}-$\tau$ correlation was fitted the exact same way as the global \ion{H}{i}-$\tau$ correlation (Sect.~\ref{sec:fit-hi-tau}).

\begin{figure}
  \centering
  \resizebox{\hsize}{!}{\includegraphics{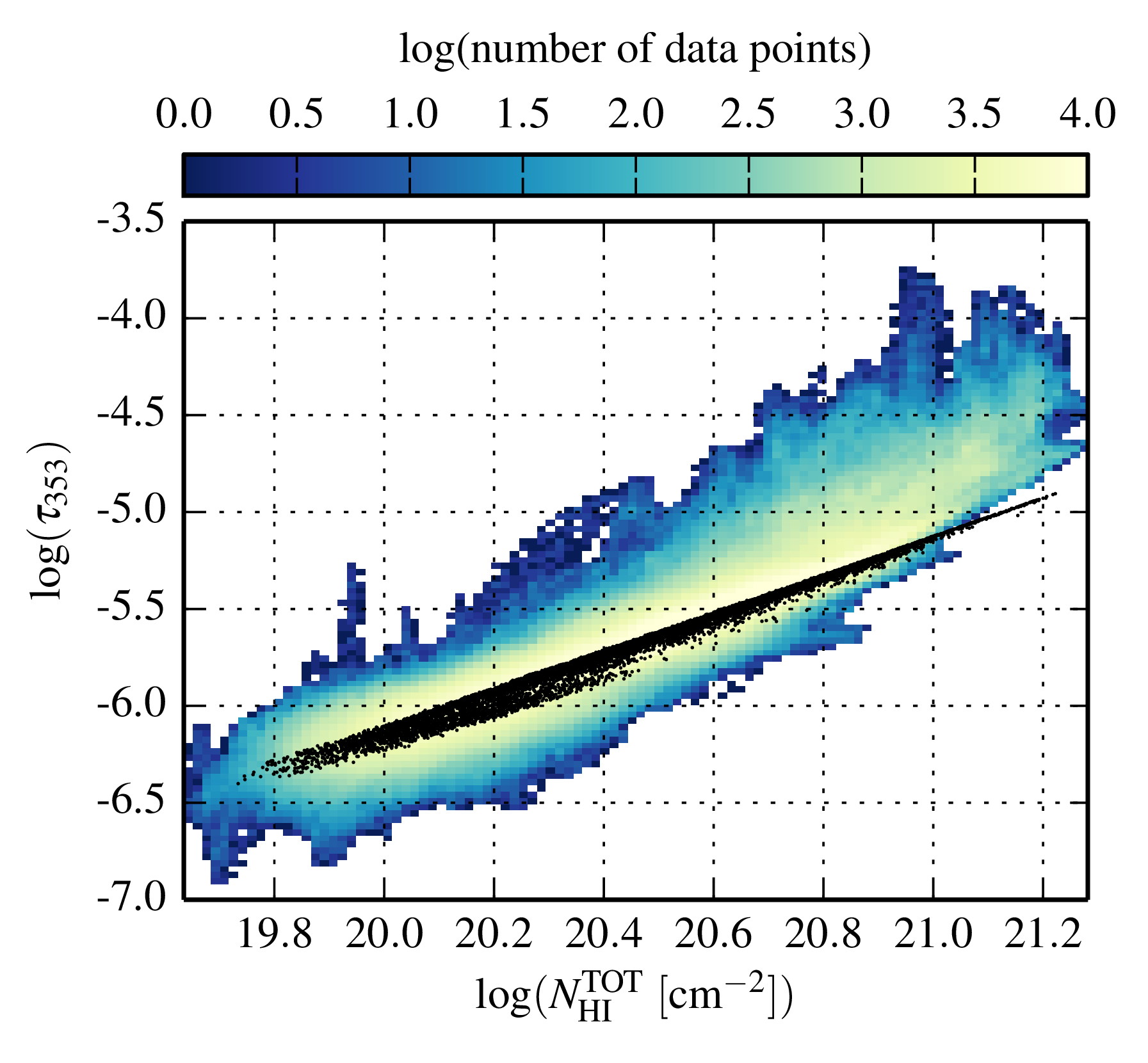}}
  \caption{\ion{H}{i}-$\tau$ correlation plot for the northern Galactic hemisphere at $b>20^\circ$. The \ion{H}{i} column density is the combined LVC and IVC emission between $-100\,\mathrm{km}\,\mathrm{s}^{-1}\leq v_\mathrm{LSR} \leq +100\,\mathrm{km}\,\mathrm{s}^{-1}$. The black dots show the two-component linear model consisting of LVC and IVC contributions. From this model 10000 data points are chosen randomly and plotted.}
  \label{fig:hi-tau-north}
\end{figure}

\section{Northern Galactic hemisphere}
\label{sec:nor-gal-hem}

\begin{figure*}
  \centering
  \resizebox{\hsize}{!}{\includegraphics[width=17cm]{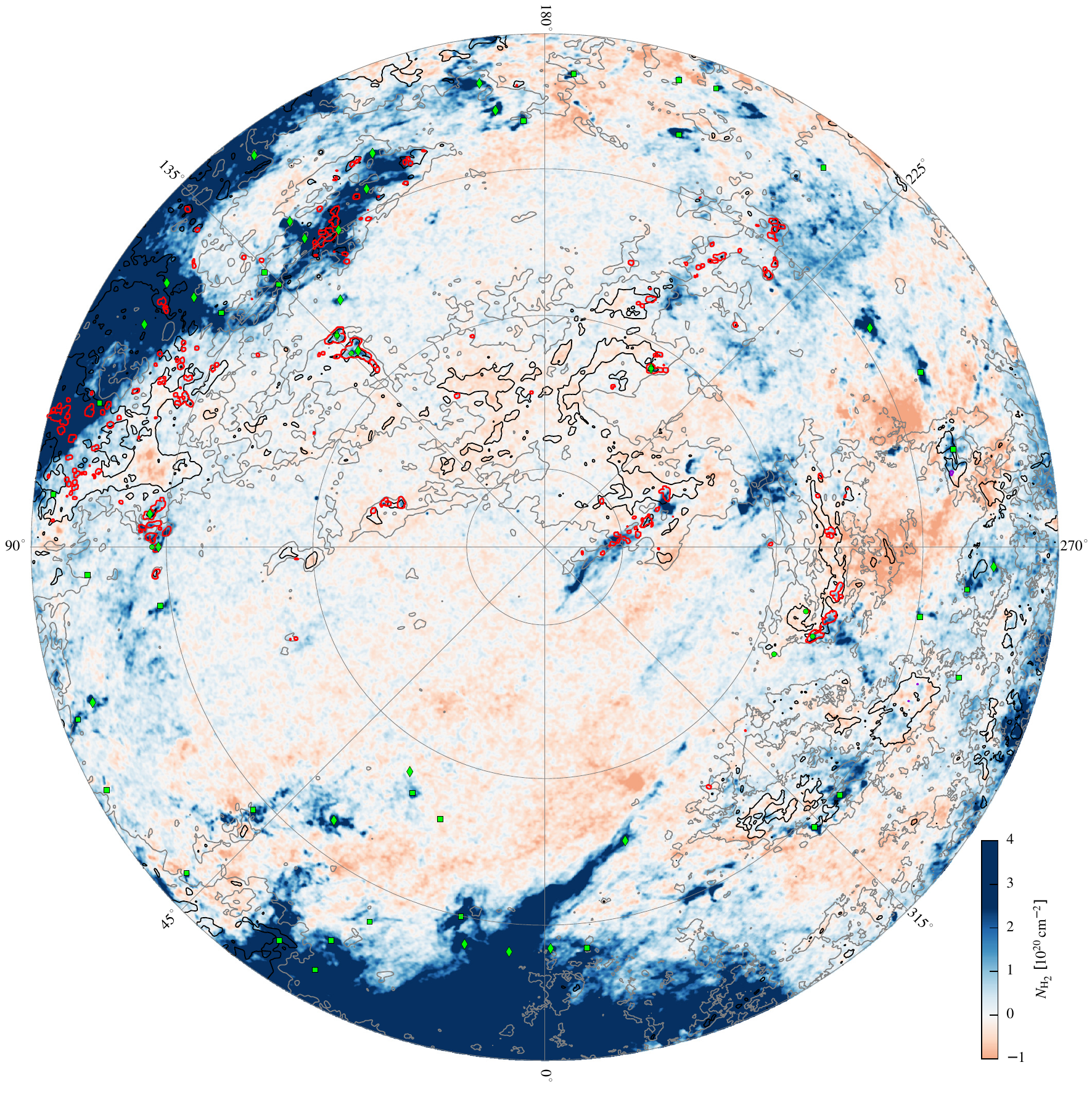}}
  \caption{Distribution of inferred H$_2$ column densities and MIVC$^\pm$ candidates for the northern Galactic hemisphere at $b>20^\circ$. The map is plotted in zenith-equal area projection centred on the northern Galactic pole with Galactic coordinates. The iso-latitude rings are placed at $20^\circ$ with an increment of $20^\circ$. Blue and red indicate positive and negative $N_{\mathrm{H}_2}$. The grey and black contours denote the total integrated (negative and positive) IVC \ion{H}{i} gas at $N_\mathrm{\ion{H}{i}}^\mathrm{IVC}=0.75\times10^{20}\,\mathrm{cm}^{-2}$ and $1.5\times10^{20}\,\mathrm{cm}^{-2}$ , respectively, tracing the location of IVC complexes. The red contours show the location of the inferred MIVC$^-$ candidates, the dark violet contours that of the MIVC$^+$ candidates (Tables~\ref{tab:mivc-candidates-large} and \ref{tab:mivc-candidates-small}). The green circles mark the location of confirmed MIVCs as listed in \citet{Magnani2010}. The green diamonds indicate known high-latitude molecular clouds from FIR excess emission, the green squares show the positions of unidentified FIR excess sources as listed in \citet[][their Tables 3 and 4]{Reach1998}.}
  \label{fig:nh2-north}
\end{figure*}

\begin{figure*}[!t]
  \centering
  \resizebox{\hsize}{!}{\includegraphics[width=17cm]{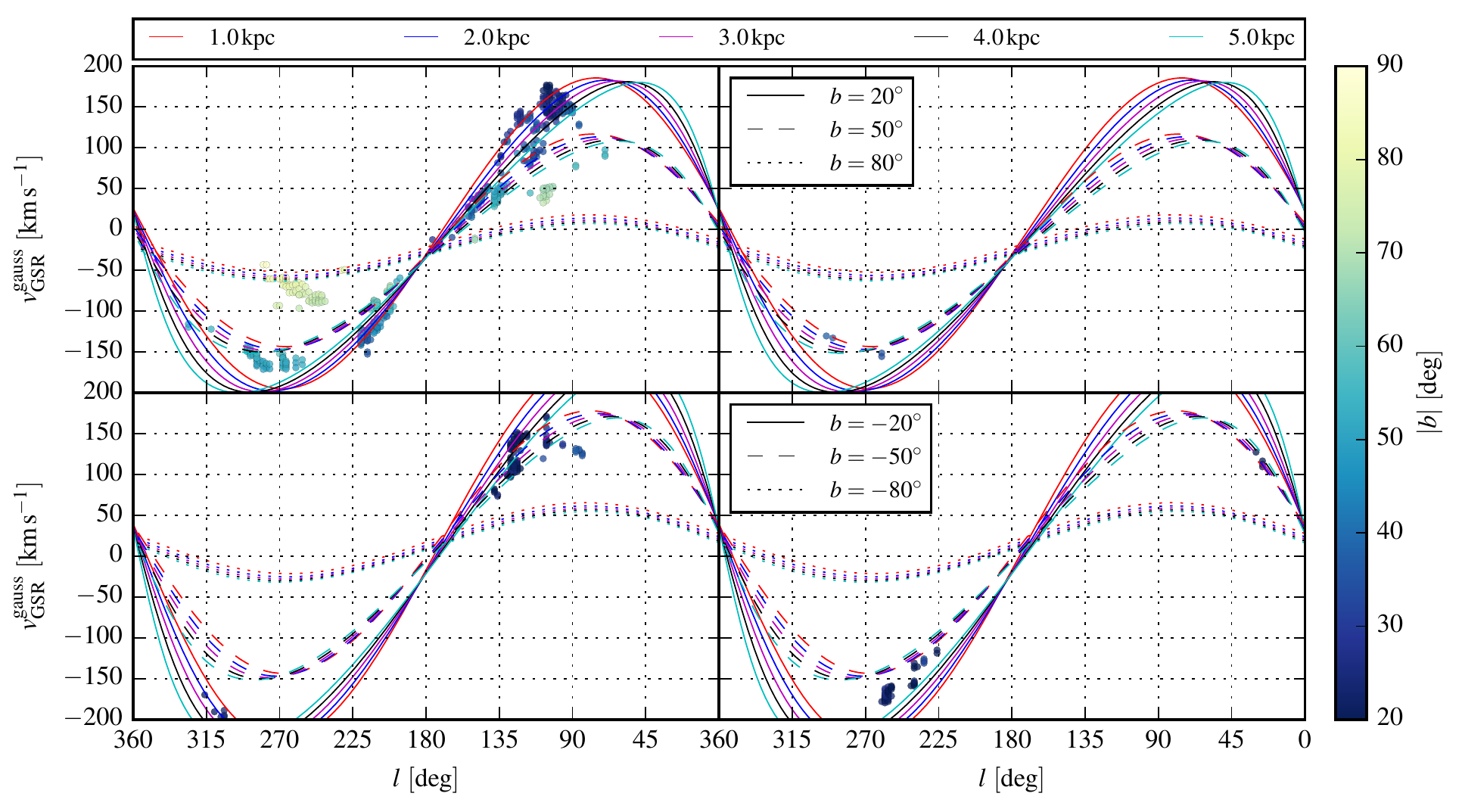}}
  \caption{Radial GSR velocities of northern (\textit{top}) and southern (\textit{bottom}) MIVC$^\pm$ samples compared with Galactic longitudes $l$, left for the MIVCs$^-$, right for the MIVCs$^+$. The velocities of the individual candidates are given by the $v_\mathrm{GSR}^\mathrm{gauss}$ from the Gaussian decomposition of the \ion{H}{i} survey data. The colour of each point encodes the absolute value of the latitude coordinate $b$ of the candidates as indicated by the colour bar. The different lines show the modelled rotation velocities including a radial, vertical, and lagging velocity component as estimated by \citet{Marasco2011}. Their vertical height $z=1.6\,$kpc is assumed, for which the velocities are modelled at some heliocentric distances $d$ and latitudes $b$ as given in the legends. The solid, dashed, and dotted lines are for $b=20^\circ$, $b=50^\circ$, and $b=80^\circ$. The red, blue, magenta, black, and cyan lines are for heliocentric distances within the Galactic plane of $d=1\,$kpc, $2\,$kpc, $3\,$kpc, $4\,$kpc, and $5\,$kpc.}
  \label{fig:kinematics}
\end{figure*}

In Fig.~\ref{fig:hi-tau-north} we plot the resulting correlation for the northern Galactic hemisphere at $b>20^\circ$. The two-component linear model is fitted and the resulting linear parameters are compiled in Table~\ref{tab:hi-tau-coeff}.

The data have a large scatter around the estimated linear model. Above $\log{(N_\mathrm{\ion{H}{i}}^\mathrm{TOT}/[\mathrm{cm}^{-2}])}\simeq20.3,$ the correlation steepens significantly, which is thought to reflect the presence of molecular hydrogen. Even at $b>20^\circ$ there are lines of sight exceeding $N_\mathrm{\ion{H}{i}}^\mathrm{TOT}\simeq10^{21}\,\mathrm{cm}^{-2}$, which indicates large amounts of molecular gas due to the large FIR excess. At such high \ion{H}{i} column densities \ion{H}{i} absorption occurs \citep[][their Fig.~10]{Strasser2004}.

The best-fit linear parameters $R$, $\epsilon^\mathrm{LVC}$, and $\epsilon^\mathrm{IVC}$ were inserted into Eq.~\eqref{eq:nh2-ivc} to calculate H$_2$ column densities across the northern Galactic hemisphere relative to the global average gas-to-dust ratios. The resulting H$_2$ distribution is shown in colour in Fig.~\ref{fig:nh2-north} with the northern Galactic pole at the centre. The grey and black contours mark the total IVC \ion{H}{i} column density at levels of $0.75$ and $1.5\times10^{20}\,\mathrm{cm}^{-2}$. Known high-latitude molecular clouds are indicated by the green diamonds \citep[][their Table 3]{Reach1998}, additional FIR-excess clouds identified by \citet[][their Table 4]{Reach1998} as green squares. The green circles show the positions of confirmed MIVCs \citep{Magnani2010}.

The new \ion{H}{i} and FIR data allow us to study the FIR excess emission on much smaller angular scales that reveal many additional excess objects. We note that the H$_2$ map presented in Fig.~\ref{fig:nh2-north} is quantitatively only valid for the IVC gas.

Generally, the IVC gas appears to be FIR deficient relative to the Galactic average. This is expected for two reasons. Firstly, IVCs typically have metallicites slightly lower than solar \citep[e.g.][]{Wakker2001}, resulting in potentially lower amounts of dust. Secondly, IVCs are located at distances of up to $\sim$2\,kpc \citep[e.g.][]{Wakker2001}, reducing the strength of the radiation field by which the dust is heated.

\subsection{Northern Galactic MIVC population}
\label{sec:mivcs-north}

At total \ion{H}{i} column densities lower than $N_\mathrm{\ion{H}{i}}^\mathrm{TOT}\leq1\times10^{20}\,\mathrm{cm}^{-2}$, we infer a median of $\tilde{N}_{\mathrm{H}_2}\simeq0.05\times10^{20}\,\mathrm{cm}^{-2}$ and a standard deviation of $\sigma(N_{\mathrm{H}_2})\simeq0.20\times10^{20}\,\mathrm{cm}^{-2}$. These quantities measure the intrinsic scatter and bias of the fitted northern \ion{H}{i}-$\tau$ correlation. At $N_\mathrm{\ion{H}{i}}^\mathrm{TOT}<1\times10^{20}\,\mathrm{cm}^{-2}$ there is no significant amount of H$_2$ expected, which is consistent with our data. The median value is significantly lower than the scatter, and at $3\sigma$ significance, a northern MIVC candidates requires $N^\mathrm{IVC}_{\mathrm{H}_2}=\tilde{N}_{\mathrm{H}_2}+3\sigma(N_{\mathrm{H}_2})\gtrsim0.65\times10^{20}\,\mathrm{cm}^{-2}$.

When we applied the four criteria for MIVCs, we identified the candidates that are indicated in Fig.~\ref{fig:nh2-north} by the red and dark violet contours for the MIVC$^-$ and MIVC$^+$ candidates, respectively. There is a very significant discrepancy in candidate numbers for the two velocity regimes towards the northern sky: Only three MIVC$^+$ candidates are detected, compared to 184 MIVC$^-$ candidates. Of the MIVC$^-$ candidates, 134 exceed the size of a single GASS beam (Table~\ref{tab:mivc-candidates-large}), 50 are smaller (Table~\ref{tab:mivc-candidates-small}). Only a single MIVC$^+$ candidate is larger (Table~\ref{tab:mivc-candidates-large}), two are smaller than the GASS beam (Table~\ref{tab:mivc-candidates-small}).

All previously known MIVCs \citep{Magnani2010} are identified as candidates. Only G\,295.0+57.1 is not recovered because it is not well separated by the LSR velocity limit at $v_\mathrm{LSR}=-20\,\mathrm{km}\,\mathrm{s}^{-1}$. Still, there is FIR excess emission associated with this object. These results confirm our definition of MIVCs (Sect.~\ref{sec:def-mivc}) and justify the simplifications made in the global \ion{H}{i}-$\tau$ correlation. The complete sample of MIVC$^\pm$ candidates is compiled in Tables \ref{tab:mivc-candidates-large} and \ref{tab:mivc-candidates-small}.

The MIVCs$^-$ are located mostly towards the outer Galactic disk at mean latitudes of ${\bar{b}\simeq45^\circ}$ and the MIVC$^+$ candidates at ${\bar{b}\simeq35^\circ}$, which is well above the limit of $b=20^\circ$. The candidates are not isolated and often part of larger coherent IVC structures, such as Draco.

\begin{figure}[!t]
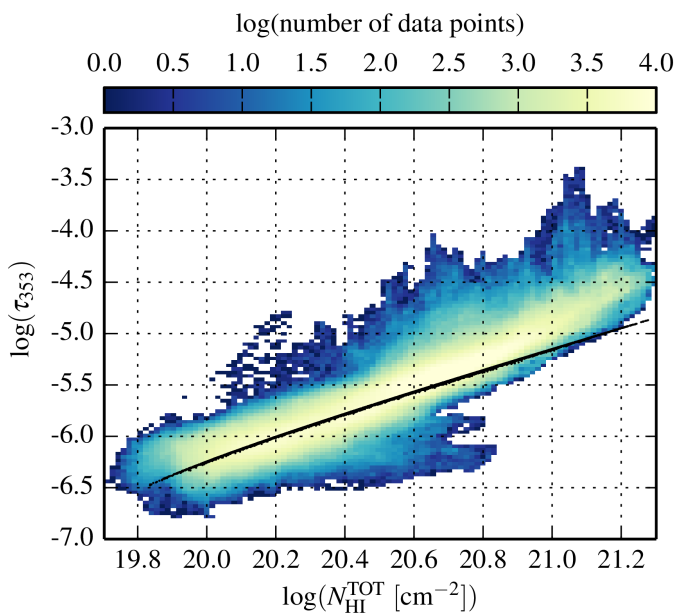

  \centering
  \includegraphics[width=.5\textwidth]{{{TOT_HI-tau353_south}}}
  \caption{\ion{H}{i}-$\tau$ correlation plot for the southern Galactic hemisphere at $b<-20^\circ$. The \ion{H}{i} column density is the combined LVC and IVC emission between $-100\,\mathrm{km}\,\mathrm{s}^{-1}\leq v_\mathrm{LSR} \leq +100\,\mathrm{km}\,\mathrm{s}^{-1}$. The black dots show the two-component linear model consisting of LVC and IVC contributions. From this model 10000 data points are chosen randomly and plotted.}
  \label{fig:hi-tau-south}
\end{figure}

\subsection{Comparison of the northern MIVC candidate sample with kinematical model}
\label{sec:kinematical-model-north}

The kinematics of the MIVC samples were compared to a model of Galactic rotation. For this we calculated the expected observable radial velocity for given coordinates $l$ and $b$ and Galactocentric radius $R$ assuming the model of Galactic rotation $\theta(R)$ as parametrised by \citet[][their Eq.~6]{Brand1993}. In this model the rotational velocity of the Sun is set to $\theta_0=220\,\mathrm{km}\,\mathrm{s}^{-1}$ and its Galactocentric distance to $R_0=8.5\,$kpc.

We included the additional radial and vertical velocity components $v_\mathrm{r}=-30^{+7}_{-5}\,\mathrm{km}\,\mathrm{s}^{-1}$ and $v_\mathrm{z}=-20^{+5}_{-7}\,\mathrm{km}\,\mathrm{s}^{-1}$ of the Galactic halo derived by \citet{Marasco2011}. \citet{Marasco2011} estimated a vertical scale height of the extra-planar gas of $h_z=1.6^{+0.6}_{-0.4}\,$kpc with a rotational vertical lagging of $-15\pm4\,\mathrm{kpc}^{-1}$, that is,~the rotational velocity decreases away from the disk. These vertical and radial velocity components were projected onto the lines of sight for $l$ and $b$ by applying Eq.~3 of \citet{Marasco2011} and converting the obtained LSR velocities into the Galactic standard of rest (GSR).

For the projection of the radial velocity component, we have to assume some height above or below the disk, for which we took $z=1.6\,$kpc as estimated by \citet{Marasco2011}. The vertical lagging of $-15\pm4\,\mathrm{km}\,\mathrm{s}^{-1}\,\mathrm{kpc}^{-1}$ was included in the rotational velocity $\theta(R,z)$.
The parametrisation of the model was changed such that it is dependent on $l$, $b$, and the heliocentric distance $d$ instead. The conversion between helio- and Galactocentric distance was calculated by applying Eq.~2 from \citet{Brand1993}. We calculated the expected radial GSR velocity from Galactic rotation, modified by vertical, radial, and lagging velocity components. We note that $d$ quantifies the heliocentric distance within the 2D plane of the Milky Way disk $z=0$.

The resulting distribution of $l$-$v_\mathrm{GSR}$ is plotted for the northern Galactic hemisphere in Fig.~\ref{fig:kinematics} (top). The plots on the left- and right-hand side in this figure show the MIVC$^-$ and MIVC$^+$ samples, respectively. The data points represent the cloud parameters, while the different lines show the modelled rotation curve for different latitudes $b$ (solid for $20^\circ$, dashed for $50^\circ$, dotted for $80^\circ$) and heliocentric distances $d$ (red for 1\,kpc, blue for 2\,kpc, magenta for 3\,kpc, black for 4\,kpc, cyan for 5\,kpc). The colours of the data points indicate the absolute value of the latitude coordinate of the candidates as shown by the colour bar.

Apparently, the distribution of the MIVC$^\pm$ candidates is well reproduced by the rotational model of \citet{Marasco2011}, either by changing Galactic latitudes or heliocentric distances. This is an argument favouring the origin of IVCs from Galactic fountains.

\begin{figure*}]
  \centering
  \resizebox{\hsize}{!}{\includegraphics[width=17cm]{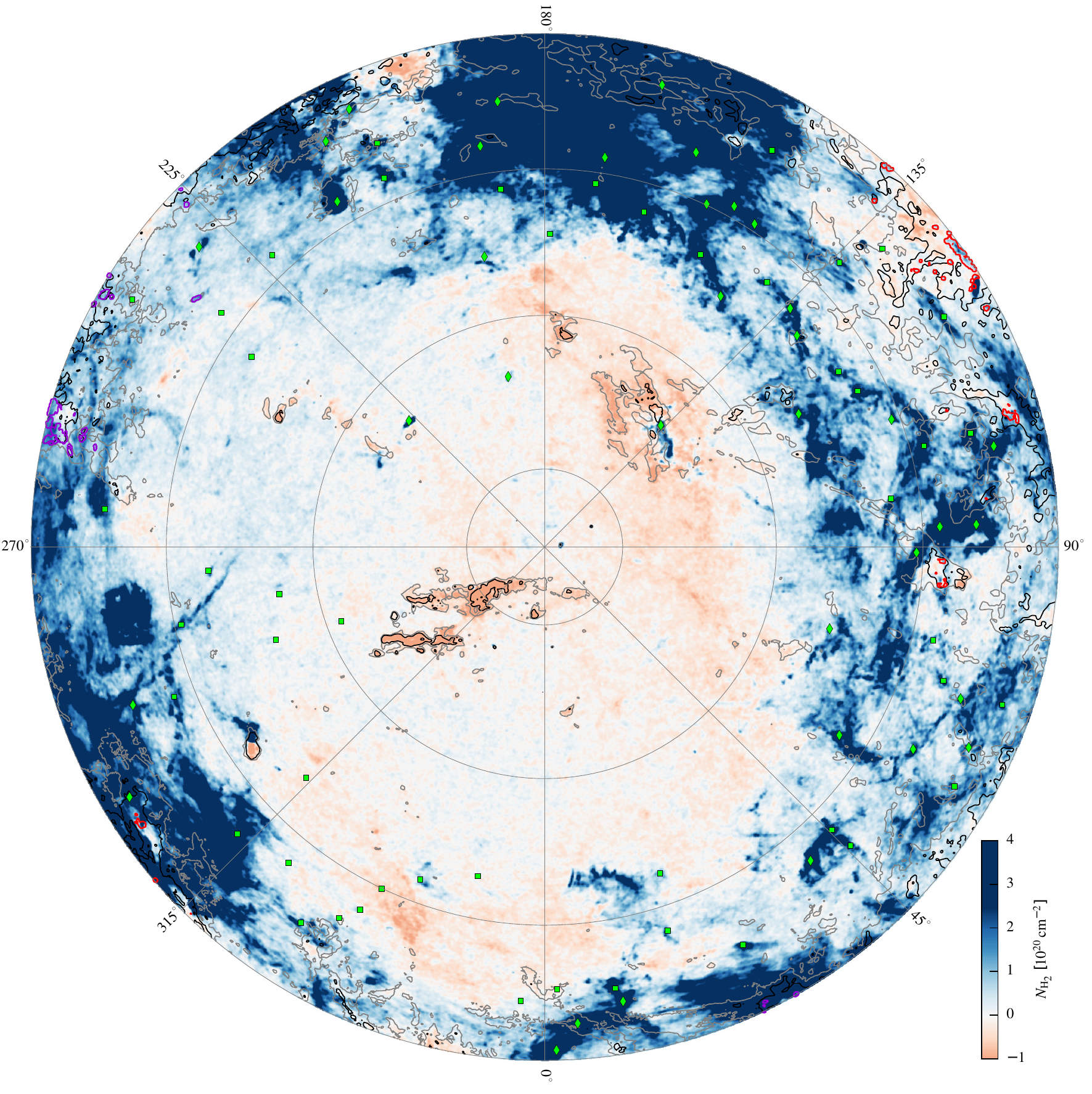}}
  \caption{Distribution of southern MIVC$^\pm$ candidates at $b<-20^\circ$ on top of the inferred H$_2$ column densities. The map is plotted in zenith-equal area projection centred on the southern Galactic pole with Galactic coordinates. The iso-latitude rings are placed at $-20^\circ$ with a decrement of $-20^\circ$. All blue areas indicate positive $N_{\mathrm{H}_2}$, all red negative $N_{\mathrm{H}_2}$. The grey and black contours denote the total integrated (negative and positive) IVC \ion{H}{i} gas at $N_\mathrm{\ion{H}{i}}^\mathrm{IVC}=0.75\times10^{20}\,\mathrm{cm}^{-2}$ and $1.5\times10^{20}\,\mathrm{cm}^{-2}$ , respectively, tracing the location of IVC complexes. The red and dark violet contours show the location of the inferred MIVC$^-$ and MIVC$^+$ candidates (Tables \ref{tab:mivc-candidates-large} and \ref{tab:mivc-candidates-small}). The green diamonds indicate known high-latitude molecular clouds from FIR excess emission, the green squares show the positions of unidentified FIR excess sources as listed in \citet[][their Tables 3 and 4]{Reach1998}.}
  \label{fig:nh2-south}
\end{figure*}

\section{Southern Galactic hemisphere}
\label{sec:sou-gal-hem}

The correlation and fitting was performed for the southern hemisphere at $b<-20^\circ$  in the same way as for the northern Galactic hemisphere (Sect.~\ref{sec:nor-gal-hem}), except for the the Magellanic System. The Magellanic System covers a large portion of the southern Galactic hemisphere that stretches over the entire LVC and IVC velocity ranges, but it is not detected in FIR emission \citep[e.g.][]{Planckcollaboration2011XXIV,Bruens2005}. Hence, the Magellanic System is expected to behave differently in terms of the \ion{H}{i}-$\tau$ correlation than the Milky Way gas, which may introduce systematic biases in the fitting of the Galactic \ion{H}{i}-$\tau$ correlation. We therefore excluded the Magellanic System when fitting the \ion{H}{i}-$\tau$ correlation. Otherwise, the Magellanic System was included in the analysis, since we cannot be sure that we masked the emission correctly.

For this masking the 3D \ion{H}{i} model of the Milky Way from \citet{Kalberla2008} was applied to separate the \ion{H}{i} emission into Galactic and Magellanic \ion{H}{i} gas. We used the same clipping values as \citet{Planckcollaboration2014XVII}, who attributed \ion{H}{i} emission to the Magellanic System when the modelled Milky Way \ion{H}{i} emission is $T_\mathrm{B}^\mathrm{model}\leq60\,$mK. This corresponds to a deviation velocity \citep{Wakker1991} of $v_\mathrm{dev}\simeq45\,\mathrm{km}\,\mathrm{s}^{-1}$.

The masking of the Magellanic System affects the estimated offset and IVC slope by a few percent. While the LVC dust emissivity $\epsilon^\mathrm{LVC}$ is comparable for both Galactic hemispheres, the IVC dust emissivity $\epsilon^\mathrm{IVC}$ is significantly higher on the southern Galactic hemisphere. The \ion{H}{i}-$\tau$ correlation plot is shown in Fig.~\ref{fig:hi-tau-south}. In
the same way as for the northern, the correlation widens at $\log{(N_\mathrm{\ion{H}{i}}^\mathrm{TOT}/[\mathrm{cm}^{-2}])}\simeq20.4,$ suggesting the presence of H$_2$ and the unusual correlation properties of the Magellanic System, which appears as very FIR dim material with $\tau\lesssim10^{-6}$. The inferred model does not produce as large a scatter as for the northern Galactic hemisphere, which reflects the dominance of the LVC gas over the IVC gas in terms of the FIR emission.

We applied the identical fitting method, except that we masked the Magellanic System. We attribute this different slope to the lack of extended high-latitude IVC gas with faint FIR emission. Instead, most of the IVC gas, as indicated by the grey and black contours, is located at lower absolute latitudes of $b>-40^\circ$ (Fig.~\ref{fig:nh2-south}), which may lead to biases of the IVC slope by the LVC material. Furthermore, the estimated offset on the southern hemisphere is negative, which is an unphysical fitting result. Potential reasons for this are discussed in Sect.~\ref{sec:disc-neg-offset}.

In the same way as for the northern Galactic hemisphere, the resulting H$_2$ distribution is plotted in Fig.~\ref{fig:nh2-south}. Towards the southern Galactic pole, some IVC gas appears to be substantially FIR deficient with respect to the Galactic average. This material is probably part of the Magellanic Stream \citep{Planckcollaboration2014XVII}.

\subsection{Southern Galactic MIVC population}
\label{sec:mivcs-south}

From the computed H$_2$ column densities at $N_\mathrm{\ion{H}{i}}^\mathrm{TOT}<1\times10^{20}\,\mathrm{cm}^{-2}$, we calculated a median $\tilde{N}_{\mathrm{H}_2}\simeq0.13\times10^{20}\,\mathrm{cm}^{-2}$ and a standard deviation $\sigma(N_{\mathrm{H}_2})\simeq0.15\times10^{20}\,\mathrm{cm}^{-2}$. This means that a southern Galactic MIVC candidate requires $\tilde{N}_{\mathrm{H}_2}+3\sigma(N_{\mathrm{H}_2})\simeq0.58\times10^{20}\,\mathrm{cm}^{-2}$ to be identified. As compared to the northern hemisphere, the offset is larger but the scatter is smaller. This reflects the smaller amount of IVC gas on the southern hemisphere.

We applied the four criteria for MIVCs and plot the identified MIVC$^-$ candidates as the red and the MIVC$^+$ candidates as the dark violet contours in Fig.~\ref{fig:nh2-south}. There are 31 MIVC$^-$ and 21 MIVC$^+$ candidates on the southern Galactic hemisphere. The 19 MIVC$^-$ candidates that are larger than the GASS beam are compiled in Table \ref{tab:mivc-candidates-large}, the 12 that are smaller in Table \ref{tab:mivc-candidates-small}. For the MIVC$^+$ candidates, 17 are larger (Table \ref{tab:mivc-candidates-large}) and four are smaller than the beam (Table \ref{tab:mivc-candidates-small}). The MIVC$^-$ candidates are located at low absolute latitudes of ${\bar{b}\simeq-23^\circ}$, the MIVC$^+$ candidates at ${\bar{b}\simeq-22^\circ}$, both are located close to the latitude limit of $b=-20^\circ$.

\subsection{Comparison of the southern MIVC candidate sample with kinematical model}
\label{sec:kinematical-model-south}

The same rotational model as for the northern Galactic hemisphere (Sect.~\ref{sec:kinematical-model-north}) was compared to the MIVC$^\pm$ candidate populations on the southern Galactic hemisphere. The resulting plots (Fig.~\ref{fig:kinematics}, bottom) show that the MIVC$^\pm$ samples are apparently consistent with the modified Galactic rotation model.

\section{Physical properties of MIVC candidates}
\label{sec:results}

Here, we estimate some basic physical properties of the northern and southern MIVC candidate ensembles. We are especially interested in the pressure and mass within these objects. For the candidates that have an angular extent smaller than the GASS beam, we calculate the respective parameters for all pixels within the beam solid angle. We explicitly distinguish between the parameters for objects that are larger or smaller than the beam.

\begin{table*}
  \caption{Median values of estimated angular extents $\theta$, spatial extents $d$, kinetic temperatures $T_\mathrm{kin}$, densities $n_\mathrm{\ion{H}{i}}$, pressures $p/k_\mathrm{B}$, molecular fractions $f_\mathrm{mol}$, \ion{H}{i} masses $M_\mathrm{\ion{H}{i}}$, H$_2$ masses $M_{\mathrm{H}_2}$, and total hydrogen masses $M_\mathrm{H}$ of northern and southern MIVC$^\pm$ samples. The samples are subdivided between angular sizes $\Omega$ larger or smaller than the GASS beam $\Omega_\mathrm{GASS}$. The quantities are computed for distances that are obtained from 1000 realisations of vertical distributions with scale heights of $h_z=0.5\,$kpc and $h_z=1.6\,$kpc, from which the distances to the candidates are calculated for their latitudes $b$ (Sect.~\ref{sec:physical-props-distances}). Note that there is only one northern MIVC$^+$ candidate with $\Omega>\Omega_\mathrm{GASS}$.}  
  \label{tab:physical-props}
  \centering
  \begin{tabular}{cccccccccc}
    \hline\hline
    & $\theta$ & $d$ & $T_\mathrm{kin}$ & $n_\mathrm{\ion{H}{i}}$ & {$p/k_\mathrm{B}$} & $f_\mathrm{mol}$ & {$M_\mathrm{\ion{H}{i}}$} & {$M_{\mathrm{H}_2}$} & {$M_\mathrm{H}$}\\
    & $[\mathrm{arcmin}]$ & $[\mathrm{pc}]$ & $[\mathrm{K}]$ & $[\mathrm{cm}^{-3}]$ & {$[\mathrm{K}\,\mathrm{cm}^{-3}]$} & & {$[\mathrm{M}_\odot]$} & {$[\mathrm{M}_\odot]$} & {$[\mathrm{M}_\odot]$}\\
    \hline
    \multicolumn{10}{c}{$\Omega\geq\Omega_\mathrm{GASS}$}\\
    \hline

    northern MIVCs$^-$ & 33.4  & \multicolumn{1}{r}{8.8/28.7} & 411 & \multicolumn{1}{r}{3.9/1.2} & \multicolumn{1}{r}{1560/500} & 0.55 & \multicolumn{1}{r}{75/760} & \multicolumn{1}{r}{90/990} & \multicolumn{1}{r}{210/2280}\\        
    southern MIVCs$^-$ & 31.4 & \multicolumn{1}{r}{13.6/42.8} & 437 & \multicolumn{1}{r}{5.2/1.8} & \multicolumn{1}{r}{2160/730} & 0.37 & \multicolumn{1}{r}{310/2480} & \multicolumn{1}{r}{140/1380} & \multicolumn{1}{r}{580/5560}\\  
    \hline
    northern MIVCs$^+$ & 23.8 & \multicolumn{1}{r}{6.9/20.0} & 472 & \multicolumn{1}{r}{14.3/4.9} & \multicolumn{1}{r}{6730/2310} & 0.28 & \multicolumn{1}{r}{90/750} & \multicolumn{1}{r}{35/290} & \multicolumn{1}{r}{120/1040} \\ 
    southern MIVCs$^+$ & 36.6 & \multicolumn{1}{r}{14.7/46.0} & 386 & \multicolumn{1}{r}{2.8/0.9} & \multicolumn{1}{r}{920/290} & 0.54 & \multicolumn{1}{r}{210/2100} & \multicolumn{1}{r}{300/3320} & \multicolumn{1}{r}{510/5280}\\ 
    \hline\hline

    \multicolumn{10}{c}{$\Omega<\Omega_\mathrm{GASS}$}\\
    \hline
    northern MIVCs$^-$ & 10.8  & \multicolumn{1}{r}{2.5/7.8} & 427 & \multicolumn{1}{r}{10.4/3.1} & \multicolumn{1}{r}{3840/1150} & 0.41 & \multicolumn{1}{r}{4.1/44} & \multicolumn{1}{r}{3.9/42} & \multicolumn{1}{r}{8.1/86}\\  
    southern MIVCs$^-$ & 9.4 & \multicolumn{1}{r}{3.1/10.6} & 482 & \multicolumn{1}{r}{15.5/4.8} & \multicolumn{1}{r}{7140/2160} & 0.28 & \multicolumn{1}{r}{15/130} & \multicolumn{1}{r}{7.7/80} & \multicolumn{1}{r}{25/240}\\  
    \hline
    northern MIVCs$^+$ & 5.2 & \multicolumn{1}{r}{1.4/4.5} & 441 & \multicolumn{1}{r}{43.9/13.3} & \multicolumn{1}{r}{19400/5960} & 0.23 & \multicolumn{1}{r}{2.7/29} & \multicolumn{1}{r}{0.8/8.9} & \multicolumn{1}{r}{3.5/37.9}\\ 
    southern MIVCs$^+$ & 10.7 & \multicolumn{1}{r}{4.4/13.8} & 510 & \multicolumn{1}{r}{16.1/5.0} & \multicolumn{1}{r}{7890/2500} & 0.34 & \multicolumn{1}{r}{25/240} & \multicolumn{1}{r}{8.0/80} & \multicolumn{1}{r}{30/310}\\ 
    \hline    
    \end{tabular}
\end{table*}

\begin{figure*}
  \centering
  \resizebox{\hsize}{!}{\includegraphics[width=17cm]{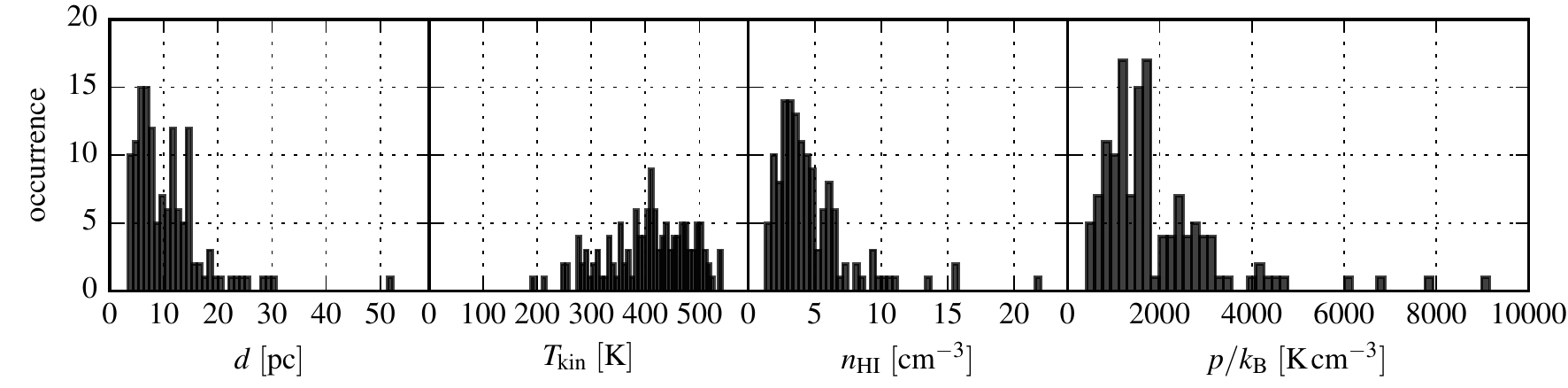}}\\
  \resizebox{\hsize}{!}{\includegraphics[width=17cm]{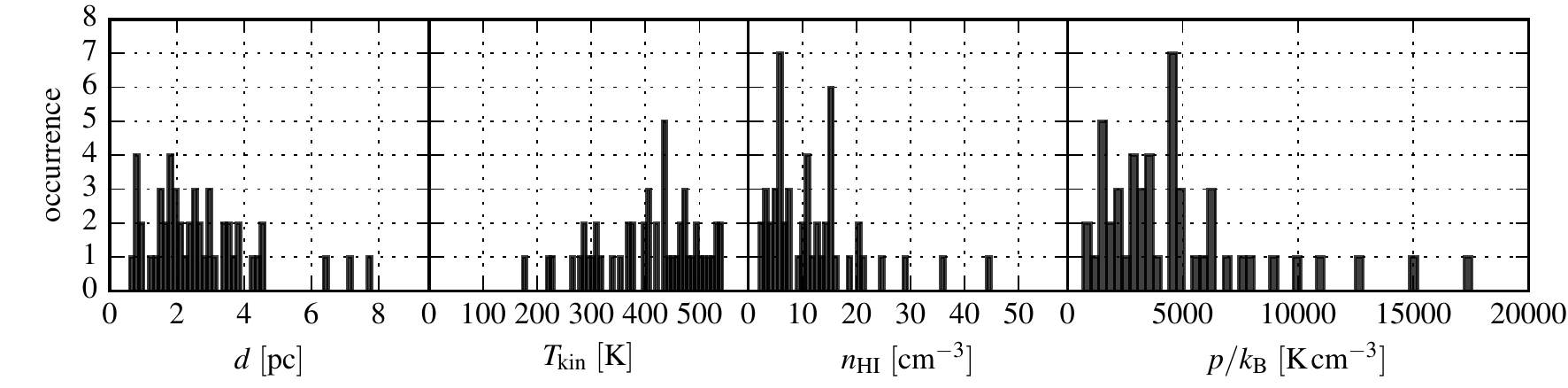}}
  \caption{Spatial sizes $d$, kinetic temperatures $T_\mathrm{kin}$, \ion{H}{i} particle densities $n_\mathrm{\ion{H}{i}}$, and pressures $p/k_\mathrm{B}$ of northern MIVC$^-$ ensemble. All objects with angular sizes exceeding the GASS beam are shown at the top, all objects smaller that are smaller at the bottom. The properties are the median values derived from 1000 realisations of the vertical distribution with a scale height of $h_z=0.5\,$kpc (Sect.~\ref{sec:physical-props-distances}).}
  \label{fig:physical-props-histo}
\end{figure*}

\subsection{Derivation}
\label{sec:physical-props-derivation}

The pressure $p$ of an \ion{H}{i} gas is calculated from the observables \ion{H}{i} particle density $n_\mathrm{\ion{H}{i}}$ and kinetic gas temperature $T_\mathrm{kin}$ by
\begin{equation}
 \label{eq:pressure}
 \frac{p}{k_\mathrm{B}} = n_\mathrm{\ion{H}{i}} T_\mathrm{kin}
\end{equation}
in units of the Boltzmann constant $k_\mathrm{B}$.

Assuming spherical symmetry, the \ion{H}{i} particle density of an individual candidate is inferred from its peak \ion{H}{i} column density and its spatial extent $d$ by
\begin{equation}
 \label{eq:density}
 n = \frac{N_\mathrm{\ion{H}{i}}}{d}.
\end{equation}
The spatial extent $d$ depends on the unknown distance $D$. The candidate clouds are most likely located within the disk-halo interface of our Galaxy, embedded in a thin and diffuse \ion{H}{i} layer \citep[e.g.][]{Kalberla2009}. This layer also contributes to the measured \ion{H}{i} column density of the candidate cloud, but its contribution is unrelated to the object. To quantify this additional contribution, we took the median \ion{H}{i} column density within $5^\circ$ around the candidate and subtracted this background column density from its peak \ion{H}{i} column density.

The angular extent of each candidate was converted into the angular diameter $\alpha$ of a sphere. The actual spatial size $d$ of a candidate depends on the distance $D$ to the cloud by $d = \alpha D$. However, the distance is generally unknown for most candidates. As discussed in Sect.~\ref{sec:physical-props-distances}, we adopted vertical distributions for the candidate samples from which the distance $D$ to each object was calculated.

The kinetic temperature of the \ion{H}{i} gas was obtained from the decomposition of \ion{H}{i} lines into Gaussian components. The observed and modelled line widths $\Delta v$ are upper limits to the kinetic gas temperature \citep[e.g.][]{Kalberla2009}:
\begin{equation}
 \label{eq:temperature}
 T_\mathrm{kin}\leq \frac{m_\mathrm{H}}{8\,k_\mathrm{B}\,\ln{2}} \left( \Delta v \right)^2 \simeq 21.8\,\mathrm{K}\left( \frac{\Delta v}{[\mathrm{km}\,\mathrm{s}^{-1}]} \right)^2
\end{equation}
with the mass of an hydrogen atom $m_\mathrm{H}$.

The \ion{H}{i} mass $M_\mathrm{\ion{H}{i}}$ of a candidate cloud is the integral of the \ion{H}{i} column densities across the angular extent $\Omega$ of the object:
\begin{equation}
 \label{eq:himass}
 M_\mathrm{\ion{H}{i}} = m_\mathrm{H} D^2 \int_\mathrm{source}\mathrm{d}\Omega\, N_\mathrm{\ion{H}{i}}(l,b).
\end{equation}
The total hydrogen mass $M_\mathrm{H}$ of a candidate cloud is the sum of the \ion{H}{i} mass $M_\mathrm{\ion{H}{i}}$ and the mass in molecular hydrogen $M_{\mathrm{H}_2}$. $M_{\mathrm{H}_2}$ is computed like Eq.~\eqref{eq:himass} for the H$_2$ column densities $N_{\mathrm{H}_2}(l,b)$ but with an additional factor of two.

The molecular fraction $f_\mathrm{mol}$ of a candidate MIVC is given by
\begin{equation}
 \label{eq:fmol}
 f_\mathrm{mol} = \frac{2 N_{\mathrm{H}_2}}{N_\mathrm{\ion{H}{i}} + 2 N_{\mathrm{H}_2}} = \frac{M_{\mathrm{H}_2}}{M_\mathrm{\ion{H}{i}} + M_{\mathrm{H}_2}}.
\end{equation}

\subsection{Distances to the IVC gas}
\label{sec:physical-props-distances}

For the well-studied MIVCs good distance constraints of $300-400\,$pc are available \citep{Gladders1998,Benjamin1996,Roehser2014}. However, for the majority of IVC gas, we lack distance constraints, except for the global upper limits of $\sim$2\,kpc \citep[e.g.][]{Wakker2001}. To obtain a rough distance estimate to the bulk IVC gas, we assumed that the IVCs are caused by a Galactic fountain process.

A Galactic fountain lifts material in the lower halo to some height $z$ described by some vertical distribution \citep[e.g.][]{Bregman1980}. We assumed that the IVC gas is distributed according to a sech$^2$ profile \citep{Marasco2011} with vertical scale heights of $h_z=0.5\,$kpc and $h_z=1.6\,$kpc. The latter height was derived by \citet{Marasco2011} for the Milky Way halo. From the height $z$, the corresponding distance $D$ is then given by $D=z/\sin{|b|}$.

Since this approach focuses on the statistical properties of the MIVC candidate samples, we calculated the parameters for 1000 different realisations of the vertical distribution. The given numbers are the median values from all realisations.

\subsection{Results}
\label{sec:physical-props-results}

In Table~\ref{tab:physical-props} we list the median of the inferred physical properties of each IVC population (northern MIVCs$^\pm$, southern MIVCs$^\pm$). {Furthermore, we distinguish between samples with angular sizes $\Omega$ larger or smaller than the GASS beam $\Omega_\mathrm{GASS}$.} For the largest sample, the northern MIVCs$^-$, we additionally show histograms of the spatial sizes $d$, kinetic temperatures $T_\mathrm{kin}$, \ion{H}{i} particle densities $n_\mathrm{\ion{H}{i}}$, and pressures $p/k_\mathrm{B}$, separately for objects larger and smaller than the beam (Fig.~\ref{fig:physical-props-histo}, top and bottom). We note that there are only three northern MIVC$^+$ candidates.

\begin{itemize}
 \item The median angular diameters of the MIVC candidates are smaller than $\sim$$37\arcmin$. This is about or below the angular resolution of the LAB survey \citep{Kalberla2005} that has been used in previous studies of the \ion{H}{i}-$\tau$ correlation. This is the reason why these studies did not resolve the candidates spatially, which rendered the candidates potentially undetectable.

  \item The median upper limits of the kinetic temperatures are $T_\mathrm{kin}\lesssim500\,$K, which is typical of the CNM \citep[e.g.][]{Kalberla2009}. The corresponding line widths are $\mathrm{FWHM}\lesssim4.8\,\mathrm{km}\,\mathrm{s}^{-1}$, which is close to the applied upper line width criterion for the definition of MIVCs of $\mathrm{FWHM}\leq5\,\mathrm{km}\,\mathrm{s}^{-1}$. We may therefore have excluded some objects as a result of this selection.

  \item The median \ion{H}{i} particle densities are $n_\mathrm{\ion{H}{i}}\simeq1-40\,\mathrm{cm}^{-3}$, typical of the CNM \citep[e.g.][]{Kalberla2009}.

  \item The median pressure is $p/k_\mathrm{B}=300-8000\,\mathrm{K}\,\mathrm{cm}^{-3}$. For the largest sample, the northern MIVCs$^-$, the median pressure within the atomic gas is $\sim$$1560\,\mathrm{K}\,\mathrm{cm}^{-3}$, which is lower than the average equilibrium thermal pressure in the ISM of $\sim$$3000\,\mathrm{K}\,\mathrm{cm}^{-3}$ \citep[e.g.][]{Wolfire2003}. Many MIVC candidates appear to be over-pressured relative to the surrounding medium when the pressure contribution from H$_2$ and the decreasing equilibrium thermal pressure with scale height \citep{Wolfire1995b} is considered.

  \item The median molecular fractions are high with $f_\mathrm{mol}\simeq0.2-0.6$, typical of diffuse and translucent molecular clouds \citep{Snow2006}.
\end{itemize}

The candidates that are smaller than the GASS beam tend to have higher Doppler temperatures than those that are more extended. In combination with their higher densities, the pressures are higher.

This may indicate unresolved velocity and angular structure within the less extended candidates. This is consistent with the results of \citet{Roehser2016}, which have been obtained from high-resolution \ion{H}{i} and CO observations of high-latitude IVCs. For the MIVC IVC\,210, they have estimate densities of up to $500\,\mathrm{cm}^{-3}$ from $^{12}$CO($1\rightarrow0$) measurements. This means that the derived parameters (Table \ref{tab:physical-props}) are certainly only upper limits for spatial extents or gas temperatures and lower limits for particle density or molecular fraction.

Furthermore, the molecular fractions of the small objects tend to be lower. This may be related to the fact that molecules form in compact unresolved structures. Our estimates are averages, which may reduce the molecular fraction systematically.

\subsection{Mass estimates}
\label{sec:physical-props-mass}

The total \ion{H}{i} and H$_2$ mass in all potential MIVC candidates was estimated from vertical sech$^2$ distributions with vertical scale heights of $h_z=0.5\,$kpc and $h_z=1.6\,$kpc for all IVCs. The latter is the height derived by \citet{Marasco2011}. Furthermore, all negative H$_2$ column densities were discarded (compare with Figs.~\ref{fig:nh2-north} and \ref{fig:nh2-south}). Again, we distinguished between objects larger or smaller than the beam.

The molecular mass constitutes a significant fraction of the total gaseous mass of the MIVC candidates (Table~\ref{tab:physical-props-mass}). Figures~\ref{fig:nh2-north} and \ref{fig:nh2-south} reveal that potential MIVCs represent only a small fraction of all IVCs. In addition, \citet{Marasco2011} found an \ion{H}{i} mass of the Milky Way halo of $3.2^{+1.0}_{-0.9}\times10^8\,\mathrm{M}_\odot$, which is significantly higher than the masses derived here. Clearly, molecular hydrogen is only a small fraction of the total gaseous mass in Galactic IVCs. This is in line with the generally low number of confirmed MIVCs and the low H$_2$ column densities in many IVCs \citep{Richter2003,Wakker2006}.

According to \citet{Marasco2011}, the upper inflow velocity of the halo material is $v_\mathrm{in}\simeq-50\,\mathrm{km}\mathrm{s}^{-1}$. This results in upper limits on the H$_2$ inflow rate of the MIVC candidates of $\dot{M}_{\mathrm{H}_2}\simeq0.006\,\mathrm{M}_\odot\,\mathrm{yr}^{-1}$ to $0.02\,\mathrm{M}_\odot\,\mathrm{yr}^{-1}$ for $h_z=0.5\,$kpc and $h_z=1.6\,$kpc. Together with the atomic mass in MIVC candidates, the inflow rate is between $\dot{M}_{\mathrm{H}}\simeq0.009\,\mathrm{M}_\odot\,\mathrm{yr}^{-1}$ and $0.03\,\mathrm{M}_\odot\,\mathrm{yr}^{-1}$.

In terms of \ion{H}{i} column density, about 2\% of the northern IVC gas with negative LSR velocity is associated with MIVC candidates. When we use this simple extrapolating factor for the atomic gas and assuming that the H$_2$ mass is negligible in the other IVC gas, the inflow from the total northern atomic and molecular IVC gas is $\sim$$0.16\,\mathrm{M}_\odot\,\mathrm{yr}^{-1}$ to $0.52\,\mathrm{M}_\odot\,\mathrm{yr}^{-1}$, which is similar to the inflow rates from Galactic HVCs \citep{Putman2012}. The estimated inflow from IVCs may contain a significant fraction of cooled halo material \citep{Marinacci2010,Marasco2012}. This combined IVC and HVC inflow rate appears to be below the Galactic star formation rate of $1.9\pm0.4\,\mathrm{M}_\odot\,\mathrm{yr}^{-1}$ \citep{Chomiuk2011}.

\setlength{\tabcolsep}{4pt}
\begin{table}
  \caption{Ensemble masses of MIVC$^\pm$ samples on northern and southern Galactic hemispheres. The values are separately calculated for objects larger or smaller than one GASS beam. The masses are median values derived from 1000 realisations of the vertical distribution with a scale height of $h_z=0.5\,$kpc (Sect.~\ref{sec:physical-props-distances}). The masses for $h_z=1.6\,$kpc \citep{Marasco2011} are about ten times higher. Only positive H$_2$ column densities are considered.}  
  \label{tab:physical-props-mass}
  \centering
  \begin{tabular}{cccc|ccc}
    \hline\hline
    & {$M_\mathrm{\ion{H}{i}}$} & {$M_{\mathrm{H}_2}$} & {$M_\mathrm{H}$} & {$M_\mathrm{\ion{H}{i}}$} & {$M_{\mathrm{H}_2}$} & {$M_\mathrm{H}$}\\
    & {$\left[\mathrm{M}_\odot\right]$} & {$\left[\mathrm{M}_\odot\right]$} & {$\left[\mathrm{M}_\odot\right]$} & {$\left[\mathrm{M}_\odot\right]$} & {$\left[\mathrm{M}_\odot\right]$} & {$\left[\mathrm{M}_\odot\right]$}\\
    \hline
    & \multicolumn{3}{c|}{$\Omega\geq\Omega_\mathrm{GASS}$} & \multicolumn{3}{c}{$\Omega<\Omega_\mathrm{GASS}$}\\
    \hline
    & \multicolumn{6}{c}{MIVC$^-$ samples}\\
    \hline
    north & 7.0$\cdot 10^4$ & 1.8$\cdot 10^5$ & 2.5$\cdot 10^5$ & 1.2$\cdot 10^3$ & 1.4$\cdot 10^3$ & 2.6$\cdot 10^3$\\
    south & 3.6$\cdot 10^4$ & 3.0$\cdot 10^4$ & 6.6$\cdot 10^4$ & 5.7$\cdot 10^2$ & 2.8$\cdot 10^2$ & 8.5$\cdot 10^2$\\
    \hline
    & \multicolumn{6}{c}{MIVC$^+$ samples}\\
    \hline
    north & 8.8$\cdot 10^1$ & 3.5$\cdot 10^1$ & 1.2$\cdot 10^2$ & 9.0$\cdot 10^0$ & 3.0$\cdot 10^0$ & 1.1$\cdot 10^1$\\
    south & 2.6$\cdot 10^4$ & 3.1$\cdot 10^4$ & 5.7$\cdot 10^4$ & 2.4$\cdot 10^2$ & 3.8$\cdot 10^2$ & 6.6$\cdot 10^2$\\
    \hline
    \end{tabular}
\end{table}
\setlength{\tabcolsep}{6pt}

\section{Discussion}
\label{sec:discussion}

We searched for potential MIVC candidates of the Milky Way towards high Galactic latitudes. All previously known MIVCs were recovered except for one single source. This positive result justifies the simplifying assumptions in the inference of MIVCs. For instance, we only considered a single gas-to-dust ratio for either northern or southern hemisphere. The fact that we still recover all the known MIVCs suggests that we captured their observational properties well. The local conditions also appear to remain substantially the same across the sky.

The complete candidate list is given in Tables \ref{tab:mivc-candidates-large} and \ref{tab:mivc-candidates-small}. These objects are ranked by analysing their \textit{\textup{local}} \ion{H}{i}-$\tau$ correlation. Only Draco and IVC\,135 are clearly associated with nearby HVC \ion{H}{i} emission, wherefore this connection has been invoked explicitly as a possible formation channel \citep{Herbstmeier1993,Moritz1998,Weiss1999,Lenz2015}. In all other cases this association is not observed. This means
that collisions between IVCs and HVCs appear not to be an important formation channel of MIVCs.

The detection of $^{12}$CO(1$\rightarrow$0) emission that is associated with the MIVC candidates is thought to be clear proof of the molecular nature of the IVCs. However, the existence of CO-dark H$_2$ gas \citep{Grenier2005,Wolfire2010,Planckcollaboration2011XIX} indicates that a stage in the transition from atomic to molecular clouds is not yet traceable by detectable amounts of CO emission \citep[also][]{Meyerdierks1996,Reach2015,DuarteCabral2016}. For this molecular component the \ion{H}{i}-FIR correlation is one of the few methods for identifying and quantifying this material.

\subsection{Selection parameters and completeness}
\label{sec:selection-completeness}

Motivated by the properties of known MIVCs, we devised selection criteria (Sect.~\ref{sec:def-mivc}) to identify the most prominent MIVCs from observations. We defined IVCs by $20\,\mathrm{km}\,\mathrm{s}^{-1}\leq|v_\mathrm{LSR}|\leq100\,\mathrm{km}\,\mathrm{s}^{-1}$ and $|b|>20^\circ$. In the following we discuss how these two criteria bias the identified population of Galactic IVCs. For this we consider a typical Galactic fountain as simulated by \citet{Melioli2008}. In their model the fountain ejects $\sim$$2.5\times10^5\,\mathrm{M}_\odot$ up to scale heights of $z\simeq2\,$kpc. The maximum downward motion (back onto the disk) is $v_\mathrm{z}\simeq-100\,\mathrm{km}\mathrm{s}^{-1}$.

\begin{figure}
  \resizebox{\hsize}{!}{\includegraphics{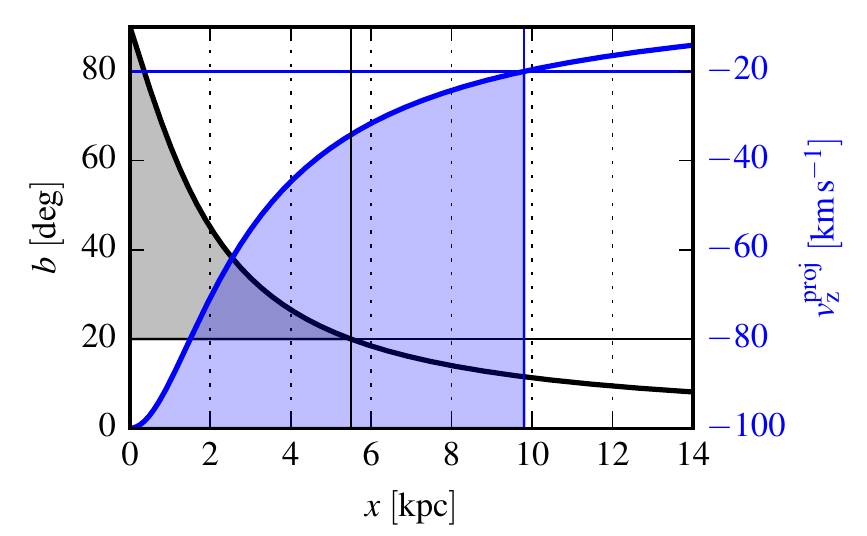}}
  \caption{Dependency of observable Galactic latitude $b$ (black) and projected vertical velocity $v_\mathrm{z}^\mathrm{proj}$ (blue) on heliocentric distance $x$ for a plane-parallel slab of fountain objects at vertical scale height $z=+2\,$kpc with a downward motion of $v_\mathrm{z}=-100\,\mathrm{km}\mathrm{s}^{-1}$. The 1D heliocentric distance $x$ is measured within the plane of the disk. The vertical and horizontal lines indicate the limits of our search for MIVCs of $20\,\mathrm{km}\mathrm{s}^{-1}\leq|v_\mathrm{LSR}|\leq100\,\mathrm{km}\mathrm{s}^{-1}$ and $|b|>20^\circ$. The coloured areas show the observable parameter spaces.}
  \label{fig:disc-selection}
\end{figure}

Matching these characteristics, we adopted a plane-parallel slab of fountain material with $|z|=2\,$kpc and $|v_\mathrm{z}|=100\,\mathrm{km}\mathrm{s}^{-1}$ down onto the disk all over the Milky Way. Because of the symmetry, both cases of $z=\pm2\,$kpc are identical. These considerations are similar to \citet{Schwarz2004}. We studied the dependency of the observable Galactic latitude coordinate $b$ and of the projected vertical velocity $v_\mathrm{z}^\mathrm{proj}$ on the heliocentric distance $x$ in one dimension (Fig.~\ref{fig:disc-selection}).
 
Analytically, at a heliocentric distance of $x\simeq5.5\,$kpc the considered fountain ejecta would be observed at $b\simeq20^\circ$ with $v_\mathrm{z}^\mathrm{proj}\simeq-34\,\mathrm{km}\mathrm{s}^{-1}$. At $x\simeq9.8\,$kpc we obtain $b\simeq12^\circ$ and $v_\mathrm{z}^\mathrm{proj}\simeq-20\,\mathrm{km}\mathrm{s}^{-1}$. For our search the global population of IVCs of the Milky Way is therefore restricted most by the latitude limit of $|b|>20^\circ$. The situation is even more severe for the MIVCs Draco, IVC\,135, and IVC\,210, which are located at $z\lesssim0.5\,$kpc and are observable only within $x\simeq1.4\,$kpc.
 
With our search parameters we trace approximately a cylindrical volume of the Milky Way with a radius of $\sim$5.5\,kpc and thickness of $2\times2\,$kpc, which is twice the height of the plane-parallel slab. We approximated the volume in our Galaxy in which star formation and fountain activity is expected to occur by the largest Galactocentric radius of high-mass star formation of $\sim$16\,kpc \citep{Reid2009}. Comparing these volumes, we find that our search may only target $\sim$12\% of the volume of interest in the Milky Way.

When we extrapolate from the local MIVC population to the entire Galaxy, this yields a global H$_2$ inflow rate from MIVCs of $\dot{M}_{\mathrm{H}_2}\simeq0.05\,\mathrm{M}_\odot\,\mathrm{yr}^{-1}-0.17\,\mathrm{M}_\odot\,\mathrm{yr}^{-1}$. Similarly, the extrapolated upper total \ion{H}{i} and H$_2$ inflow rate from IVCs is $\dot{M}_{\mathrm{H}}\simeq1.3\,\mathrm{M}_\odot\,\mathrm{yr}^{-1}-4.3\,\mathrm{M}_\odot\,\mathrm{yr}^{-1}$. Despite the considerable uncertainties in these estimates, we can draw two main conclusions from this result.
\begin{enumerate}
 \item The total in-fall of atomic and molecular material in the form of IVC gas onto the Milky Way Galaxy as a whole may account for the entire inflow that is required to sustain the current Galactic star formation rate.
 \item A vertical scale height of $h_z\lesssim0.5\,$kpc appears to be more plausible for the class of MIVCs. Larger heights lead to unrealistically high inflow rates, even more so since there are additional fuelling sources \citep{Putman2012}. This also supports the idea that MIVCs constitute a late stage in the Galactic fountain cycle. Thus, IVCs may turn molecular when they reach the disk-halo interface during their descent. Here, the environmental pressure is sufficiently high to form high density  and high column density environments, allowing the rapid formation of molecules and the protection against dissociating radiation \citep[][their Fig.~14]{Roehser2014}.
\end{enumerate}

\subsection{Comparison with \texorpdfstring{\citet{Reach1998}}{Reach et al. (1998)}}
\label{sec:disc-comp-reach1998}

The last all-sky study of the FIR excess emission was performed by \citet{Reach1998}. Today, we analyse significantly improved data in \ion{H}{i} and the FIR. This is one of the main reasons why we identify many more FIR excess objects.

The most important difference between our analysis and that of \citet{Reach1998} is that they fitted a single-component model integrated over $-100\,\mathrm{km}\,\mathrm{s}^{-1}\leq v_\mathrm{LSR} \leq +100\,\mathrm{km}\,\mathrm{s}^{-1}$. This means that they did not consider the radial velocity information. Furthermore, \citeauthor{Reach1998} fitted the \ion{H}{i}-FIR correlation in cells with radius of $10^\circ$ on a regular grid with $10^\circ$ spacing. Towards some fields this subdivision does not allow a robust estimation of their linear parameters, for example because of large molecular cloud complexes. It is therefore worthwhile to compare our results more closely to those of \citet{Reach1998}.

In Sects.~\ref{sec:mivcs-north} and \ref{sec:mivcs-south} we quantified the statistical significance of the inferred FIR excess. Using these values on the northern Galactic hemisphere at $b>20^\circ$, we found 694 statistically significant FIR excess objects that have angular sizes larger than a GASS beam and 232 objects that are larger than an LAB beam of $36\,\arcmin$ \citep{Kalberla2005}. The LAB data comprise the Leiden-Dwingeloo \ion{H}{i} survey \citep{Hartmann1997}, which were used by \citet{Reach1998} as well. In comparison to the 232~objects that are reported here, \citet{Reach1998} identified 56 FIR excess objects on the northern Galactic hemisphere.

For $b<-20^\circ$, we identified 389 statistically significant FIR excess objects that are larger than a GASS beam and 105 objects larger than a LAB beam, compared to 85 FIR excess objects reported by \citet{Reach1998}.

\citeauthor{Reach1998} reported thresholds for their FIR excess derived at $100\,\mu\mathrm{m}$ of 1.0\,MJy\,$\mathrm{sr}^{-1}$ at $|b|>20^\circ$ and of 0.3\,MJy\,$\mathrm{sr}^{-1}$ at $|b|>45^\circ$ (compare with their Fig.~6). These thresholds measure the intrinsic scatter of their residual FIR emission. To quantify similar thresholds, we performed our fitting of the \ion{H}{i}-FIR correlation data as before, but with the IRIS $100\,\mu\mathrm{m}$ data \citep{MivilleDeschenes2005b}. We find that the positive FIR excess dominates above $\sim$0.3\,MJy\,$\mathrm{sr}^{-1}$ for $|b|>20^\circ$ and above $\sim$0.2\,MJy\,$\mathrm{sr}^{-1}$ for $|b|>45^\circ$. This shows that our methods and our data allow us to deduce more reliable results that appear to be less affected by intrinsic variations of the \ion{H}{i}-FIR correlation than those of \citet{Reach1998}.

\subsection{Differences between Galactic hemispheres}
\label{sec:disc-diff-hemis}

There are some notable differences between the populations of MIVC candidates on the northern and southern Galactic hemispheres, both in terms of number and distribution. On the northern hemisphere there are 184 MIVC$^-$ and only three MIVC$^+$ candidates, on the southern there are 31 MIVC$^-$ and 21 MIVC$^+$ candidates.

If the Galactic IVC population is connected to a Galactic fountain process \citep[e.g.][]{Putman2012}, then it is unlikely to find such a significant difference in candidate numbers between the
two hemispheres. The lack of high-latitude southern IVC structures has been known for a long time now \citep[e.g.][]{Wakker2004}, but there is still no explanation.

There is a clear lack of northern MIVC$^+$ candidates. To some degree this difference is related to the different amounts of northern IVC$^\pm$ gas: At $b>20^\circ$ there is about three times more total \ion{H}{i} column density in IVCs$^-$ than in IVCs$^+$. Furthermore, the median ratio $N_\mathrm{\ion{H}{i}}^{\mathrm{IVC}^-}/N_\mathrm{\ion{H}{i}}^\mathrm{LVC}$ is about four times higher than $N_\mathrm{\ion{H}{i}}^{\mathrm{IVC}^+}/N_\mathrm{\ion{H}{i}}^\mathrm{LVC}$ at $b>20^\circ$. Combined, this may account for a factor of $\sim$10 in candidate numbers. Still, this is not sufficient to explain the lack of northern MIVC$^+$ candidates since at least ten northern MIVC$^+$ candidates would be expected.

We suggest that the lack of northern MIVC$^+$ candidates is not only related to observational biases due to the different amounts of IVC$^-$ and IVC$^+$ gas and their distribution relative to the LVC gas. Instead, there may be physical reasons for the formation and existence of MIVCs in general that are linked to the kinematics and direction of motion of the IVCs within the Galactic fountain cycle, as argued by \citet{Roehser2014,Roehser2016}. Only inflowing objects, for instance, objects~moving towards the Galactic disk, appear to form large amounts of molecular gas that are traceable by the \ion{H}{i}-FIR correlation.

The distributions of the northern and southern MIVC candidates are also clearly distinct. In the north, the candidates are located preferentially at high Galactic latitudes well above the latitude limit of $b>20^\circ$. In the south, the candidates are found at $b>-30^\circ$. Some southern IVC structures coherently connect to structures that extend up to the Galactic disk. This suggests that at least some of the southern IVC structures are objects of the disk and not the halo. Furthermore, at lower absolute latitudes there is typically more LVC gas. This hinders the evaluation of the southern candidates.

\subsection{Negative offset on the southern Galactic hemisphere}
\label{sec:disc-neg-offset}

The \ion{H}{i}-$\tau$ correlations on the northern and southern Galactic hemispheres were fitted  in the exact same way. However,
in the south, a negative offset in the FIR model is obtained (Table \ref{tab:hi-tau-coeff}). Such a negative offset is unphysical since there is no negative FIR emission.

The only difference in method is the masking of the Magellanic System on the southern hemisphere. This masking, in fact, slightly
alleviates the problem. This offset indicates that the modelled \ion{H}{i}-$\tau$ correlation is too steep, which suggests some bias that is introduced by the distribution of gas and dust on the southern Galactic hemisphere.

This offset might be related to the spatial offset of the Sun above the Galactic plane of $20.5\pm3.5\,$pc \citep{Humphreys1995}. This height is a significant fraction of that of the molecular and cold atomic gas in our Galaxy \citep[e.g.][his Table 3]{Kalberla2003},
which means that there should be more cold and dense gas towards the southern than towards the northern Galactic hemisphere. This is consistent with our \ion{H}{i} data. The total \ion{H}{i} column density at $b<-20^\circ$ is $\sim$23\% higher than at $b>20^\circ$.

Furthermore, the mid-plane pressure and the amount of molecular gas in the disk of galaxies are correlated \citep{Wong2002,MacLow2012}. This suggests that in the innermost plane, where the vertical pressure is highest, the atomic hydrogen is most efficiently converted into molecular hydrogen. There may therefore be more molecular gas at lower \ion{H}{i} column densities on the southern Galactic hemisphere, which causes the slight bias in the southern \ion{H}{i}-$\tau$ correlation.

\section{Summary}
\label{sec:summary}

We studied the IVC population of the Milky Way and its FIR properties by correlating the \ion{H}{i} and FIR emission at Galactic latitudes $|b|>20^\circ$. The recent \ion{H}{i} single-dish surveys EBHIS and GASS were complemented by the dust model obtained from \textit{Planck} FIR data. This new data set allows an exceptional analysis of global ISM properties of the Milky Way. 

The \ion{H}{i} and FIR data are correlated at the angular resolution of GASS of $\theta\simeq16.1\arcmin$. By subtracting the FIR emission associated with the atomic gas, we derived H$_2$ column densities across the sky with the focus on the molecular content of IVCs and on the search for molecular IVCs (MIVCs). We defined MIVCs from observables obtained from the \ion{H}{i} and FIR data only. 

The all-sky H$_2$ maps (Figs.~\ref{fig:nh2-north} and \ref{fig:nh2-south}) reveal a multitude of small FIR excess objects that are not apparent in previous studies of the large-scale FIR emission and excess \citep[e.g.][]{Reach1998}. This stresses the wealth of information obtained by the new all-sky survey data.

We retrieved all previously known MIVCs except for a single source, which is not identified because of our velocity selection between LVC and IVC gas. In total, we identified 239 MIVC candidates on the two Galactic hemispheres for both negative and positive radial velocities as compiled in Tables \ref{tab:mivc-candidates-large} and \ref{tab:mivc-candidates-small}. If these candidates are real MIVCs, then we should be able to detect associated $^{12}$CO(1$\rightarrow$0) emission, although during the initial stages of formation CO-dark gas may dominate the H$_2$ content of a molecular cloud \citep{Wolfire2010}. The existing CO large-scale surveys fail to detect these objects.

The numbers and distributions of candidates differ strongly between the two hemispheres and between the two radial velocity regimes: There are many more candidates found on the northern hemisphere with negative radial velocities. Such a clear dichotomy between north and south is unexpected when we assume that the IVC gas is related to a Galactic fountain process. The lack of MIVC candidates with positive radial velocities appears not to be accounted for by a different amount and distribution of LVC and IVC gas. Instead, the formation of MIVCs may be related to physical mechanisms that are connected to the inflow of the clouds onto the Galactic disk. The association of HVCs with MIVC candidates is only valid for a few objects, most notably for Draco and IVC\,135, for which a possible interaction between the MIVCs and HVCs has been explicitly mentioned before \citep{Herbstmeier1993,Moritz1998,Weiss1999,Lenz2015}.

The MIVC samples appear to be compatible with a model of Galactic rotation that contains radial, vertical, and lagging velocity components for different scale heights and latitudes. This may suggest that the sample is associated with Galactic rotation in the disk and that some fraction of these objects is very likely connected to Galactic fountain processes within the Galactic disk.

The estimated maximum inflow rate derived from our analysis of IVCs in the local Galactic environment is $\sim$$0.52\,\mathrm{M}_\odot\,\mathrm{yr}^{-1}$. We applied a latitude limit of $|b|>20^\circ$, which allowed
us to probe only $\sim$10\% of the volume of the Milky Way, in which IVCs are expected to be located. Extrapolating from the local IVC population to the entire Milky Way, the in-falling atomic and molecular IVC gas may account for the main fraction of the matter inflow onto our Galaxy. We derived plausible IVC inflow rates for vertical distributions with scale heights of $h_z\lesssim0.5\,$kpc. Larger heights exceed the expected mass inflow by far, suggesting that the class of MIVCs constitutes a late stage during the descent in the Galactic fountain cycle.

Future studies of the Galactic IVC population should focus on lower latitudes, for which, however, the \ion{H}{i}-FIR correlation becomes increasing uncertain because many emission components
are mixed along the lines-of-sight. Sophisticated modelling approaches are required to extend the traceable volume so that a more complete picture of the Galactic fountain and halo material in the Milky Way can be obtained.

\begin{acknowledgements}
The authors thank the Deutsche Forschungsgemeinschaft (DFG) for financial support under the research grant KE757/11-1. The work is based on observations with the 100$\,$m telescope of the MPIfR (Max-Planck-Institut f{\"u}r Radioastronomie) at Effelsberg. The Planck satellite is operated by the European Space Agency. The development of Planck has been supported by: ESA; CNES and CNRS/INSU-IN2P3-INP (France); ASI, CNR, and INAF (Italy); NASA and DoE (USA); STFC and UKSA (UK); CSIC, MICINN and JA (Spain); Tekes, AoF and CSC (Finland); DLR and MPG (Germany); CSA (Canada); DTU Space (Denmark); SER/SSO (Switzerland); RCN (Norway); SFI (Ireland); FCT/MCTES (Portugal); and PRACE (EU). Some figures have been prepared with the Kapteyn package \citep{KapteynPackage}. T.~R.~is a member of the International Max Planck Research School (IMPRS) for Astronomy and Astrophysics at the Universities of Bonn and Cologne as well as of the Bonn-Cologne Graduate School of Physics and Astronomy (BCGS). 
\end{acknowledgements}

\footnotesize
\begin{longtab}
\begin{longtable}{crcrcccrrc}
\caption{Locations of MIVC$^\pm$ candidates with angular sizes $\Omega$ larger than one GASS beam. The columns give the Galactic longitude $l$ and latitude $b$, the radial velocity $v_\mathrm{LSR}$, the fitted IVC dust emissivity $\epsilon^\mathrm{IVC}$ and its error, the peak H$_2$ column density $N_{\mathrm{H}_2}^\mathrm{peak}$ and its error, the line width $\mathrm{FWHM}$, the angular size $\Omega$, the total \ion{H}{i} flux $S_\mathrm{tot}$ within $-100\,\mathrm{km}\,\mathrm{s}^{-1}\leq v_\mathrm{LSR} \leq -20\,\mathrm{km}\,\mathrm{s}^{-1}$ or $+20\,\mathrm{km}\,\mathrm{s}^{-1}\leq v_\mathrm{LSR} \leq +100\,\mathrm{km}\,\mathrm{s}^{-1}$, the rank for the samples with negative or positive LSR velocities as determined by the significance $S$ (Eq.~\ref{eq:significance}), and the MIVC designation if it is a known cloud. If there are spectral components towards a single candidate that are separated by more than $10\,\mathrm{km}\,\mathrm{s}^{-1}$, then we give these two different spectral components with individual centre velocities and line widths. If $\epsilon^\mathrm{IVC}$ and $N_{\mathrm{H}_2}^\mathrm{peak}$ are both $0.00\pm0.00$, then the candidate is not fitted because either the LVC and IVC \ion{H}{i} column densities within $2^\circ\times2^\circ$ around the candidate are higher than $4\times10^{20}\,\mathrm{cm}^{-2}$.}\\
\hline\hline
   {$l$} & \multicolumn{1}{c}{$b$} & {$v_\mathrm{LSR}$} & \multicolumn{1}{c}{{$\epsilon^\mathrm{IVC}$}} & {$N^\mathrm{peak}_{\mathrm{H}_2}$} & $\mathrm{FWHM}$ & $\Omega$ & \multicolumn{1}{c}{$S_\mathrm{tot}$} & {no.} & {designation}\\
   {$[\mathrm{deg}]$} & {$[\mathrm{deg}]$} & {$[\mathrm{km}\,\mathrm{s}^{-1}]$} & \multicolumn{1}{c}{{$[10^{-26}\,\mathrm{cm}^{2}]$}} & {$[10^{20}\,\mathrm{cm}^{-2}]$} & {$[\mathrm{km}\,\mathrm{s}^{-1}]$} & {[$\mathrm{deg}^2$]} & {[Jy\,km\,s$^{-1}$]} & & \\
\hline
\endfirsthead
\caption{continued.}\\
\hline\hline
   {$l$} & \multicolumn{1}{c}{$b$} & {$v_\mathrm{LSR}$} & \multicolumn{1}{c}{{$\epsilon^\mathrm{IVC}$}} & {$N^\mathrm{peak}_{\mathrm{H}_2}$} & $\mathrm{FWHM}$ & $\Omega$ & \multicolumn{1}{c}{$S_\mathrm{tot}$} & {no.} & {designation}\\
   {$[\mathrm{deg}]$} & {$[\mathrm{deg}]$} & {$[\mathrm{km}\,\mathrm{s}^{-1}]$} & \multicolumn{1}{c}{{$[10^{-26}\,\mathrm{cm}^{2}]$}} & {$[10^{20}\,\mathrm{cm}^{-2}]$} & {$[\mathrm{km}\,\mathrm{s}^{-1}]$} & {[$\mathrm{deg}^2$]} & {[Jy\,km\,s$^{-1}$]} & & \\
\hline
\endhead
\hline
\endfoot

\multicolumn{10}{c}{$v_\mathrm{LSR}<-20\,\mathrm{km}\,\mathrm{s}^{-1}$}                                                                                                                                                 \\                      
                                                                                                                                                                                
\hline                                                                                                                                                                          
                                                                                                                                                                                        
69.8    &       55.5    &       -25.8   &       0.36    $\pm$   0.02    &       0.68    $\pm$   0.21    &       3.7     &       0.11    &       114     &       108     &               \\
84.5    &       -36.6   &       -53.1   &       0.25    $\pm$   0.02    &       2.61    $\pm$   0.18    &       4.8     &       0.12    &       272     &       123     &               \\
84.5    &       -37.5   &       -46.5   &       -2.20   $\pm$   0.30    &       0.10    $\pm$   0.02    &       4.6     &       0.11    &       262     &       150     &               \\
86.0    &       38.4    &       -44.5   &       0.75    $\pm$   0.01    &       0.35    $\pm$   0.10    &       3.8     &       0.72    &       982     &       11      &               \\
87.9    &       -37.3   &       -45.2   &       -0.01   $\pm$   0.02    &       30.01   $\pm$   12.57   &       4.7     &       0.30    &       710     &       145     &               \\
89.8    &       38.7    &       -25.2   &       2.44    $\pm$   0.05    &       1.66    $\pm$   0.06    &       3.8     &       0.60    &       606     &       22      &       Draco   \\
\end{longtable}
\end{longtab}

\footnotesize
\begin{longtab}
\begin{longtable}{crcccccrrc}
\caption{Locations of MIVC$^\pm$ candidates with angular sizes $\Omega$ smaller than one GASS beam. The columns give the Galactic longitude $l$ and latitude $b$, the radial velocity $v_\mathrm{LSR}$, the fitted IVC dust emissivity $\epsilon^\mathrm{IVC}$ and its error, the peak H$_2$ column density $N_{\mathrm{H}_2}^\mathrm{peak}$ and its error, the line width $\mathrm{FWHM}$, the angular size $\Omega$, the total \ion{H}{i} flux $S_\mathrm{tot}$ within $-100\,\mathrm{km}\,\mathrm{s}^{-1}\leq v_\mathrm{LSR} \leq -20\,\mathrm{km}\,\mathrm{s}^{-1}$ or $+20\,\mathrm{km}\,\mathrm{s}^{-1}\leq v_\mathrm{LSR} \leq +100\,\mathrm{km}\,\mathrm{s}^{-1}$, the rank for the samples with negative or positive LSR velocities as determined by the significance $S$ (Eq.~\ref{eq:significance}), and the MIVC designation if it is a known cloud. If there are spectral components towards a single candidate that are separated by more than $10\,\mathrm{km}\,\mathrm{s}^{-1}$, then we give these two different spectral components with individual centre velocities and line widths. If $\epsilon^\mathrm{IVC}$ and $N_{\mathrm{H}_2}^\mathrm{peak}$ are both $0.00\pm0.00$, then the candidate is not fitted because either the LVC and IVC \ion{H}{i} column densities within $2^\circ\times2^\circ$ around the candidate are higher than $4\times10^{20}\,\mathrm{cm}^{-2}$.}\\
\hline\hline
   {$l$} & \multicolumn{1}{c}{$b$} & {$v_\mathrm{LSR}$} & \multicolumn{1}{c}{{$\epsilon^\mathrm{IVC}$}} & {$N^\mathrm{peak}_{\mathrm{H}_2}$} & $\mathrm{FWHM}$ & $\Omega$ & \multicolumn{1}{c}{$S_\mathrm{tot}$} & {no.} & {designation}\\
   {$[\mathrm{deg}]$} & {$[\mathrm{deg}]$} & {$[\mathrm{km}\,\mathrm{s}^{-1}]$} & \multicolumn{1}{c}{{$[10^{-26}\,\mathrm{cm}^{2}]$}} & {$[10^{20}\,\mathrm{cm}^{-2}]$} & {$[\mathrm{km}\,\mathrm{s}^{-1}]$} & {[$\mathrm{deg}^2$]} & {[Jy\,km\,s$^{-1}$]} & & \\
\hline
\endfirsthead
\caption{continued.}\\
\hline\hline
   {$l$} & \multicolumn{1}{c}{$b$} & {$v_\mathrm{LSR}$} & \multicolumn{1}{c}{{$\epsilon^\mathrm{IVC}$}} & {$N^\mathrm{peak}_{\mathrm{H}_2}$} & $\mathrm{FWHM}$ & $\Omega$ & \multicolumn{1}{c}{$S_\mathrm{tot}$} & {no.} & {designation}\\
   {$[\mathrm{deg}]$} & {$[\mathrm{deg}]$} & {$[\mathrm{km}\,\mathrm{s}^{-1}]$} & \multicolumn{1}{c}{{$[10^{-26}\,\mathrm{cm}^{2}]$}} & {$[10^{20}\,\mathrm{cm}^{-2}]$} & {$[\mathrm{km}\,\mathrm{s}^{-1}]$} & {[$\mathrm{deg}^2$]} & {[Jy\,km\,s$^{-1}$]} & & \\
\hline
\endhead
\hline
\endfoot

\multicolumn{10}{c}{$v_\mathrm{LSR}<-20\,\mathrm{km}\,\mathrm{s}^{-1}$}                                                                                                                                                 \\                      
                                                                                                                                                                                
\hline                                                                                                                                                  
                                                                                                                                                                                        
70.0    &       54.8    &       -22.7   &       0.58    $\pm$   0.02    &       0.42    $\pm$   0.13    &       3.8     &       0.012   &       12.2    &       28      &               \\
84.3    &       -37.2   &       -46.4   &       0.98    $\pm$   0.21    &       0.96    $\pm$   0.06    &       4.6     &       0.003   &       16.0    &       55      &               \\
85.2    &       -36.8   &       -48.8   &       0.13    $\pm$   0.02    &       4.78    $\pm$   0.28    &       4.6     &       0.056   &       25.8    &       49      &               \\
86.1    &       -38.1   &       -42.5   &       0.29    $\pm$   0.01    &       4.24    $\pm$   0.18    &       5.0     &       0.018   &       16.3    &       23      &               \\
87.2    &       57.8    &       -38.9   &       0.21    $\pm$   0.01    &       1.21    $\pm$   0.31    &       4.3     &       0.041   &       5.0     &       17      &               \\
93.1    &       23.2    &       -35.6   &       0.81    $\pm$   0.02    &       0.59    $\pm$   0.09    &       4.7     &       0.068   &       12.7    &       7       &               \\
\end{longtable}
\end{longtab}

%
%

\bibliographystyle{aa} 
\bibliography{Literatur} 

\end{document}